\documentclass[ aps,
                pra,
                showpacs,
                amssymb,
                nofootinbib,
                superscriptaddress,
                twocolumn,
                longbibliography]{revtex4-1}
\usepackage[ansinew]{inputenc}
\usepackage{bbm}
\usepackage{bm}
\usepackage{amsbsy}
\usepackage{amsthm}
\usepackage{amssymb}
\usepackage{amsfonts}
\usepackage{amsmath, mathtools}
\usepackage{dsfont}
\usepackage{graphicx}
\usepackage{epsfig}
\usepackage{epstopdf}
\usepackage{dsfont}
\usepackage{multibib}
\usepackage{xcolor}
\usepackage[colorlinks]{hyperref}
\makeatletter
\newcommand\org@hypertarget{}
\let\org@hypertarget\hypertarget
\renewcommand\hypertarget[2]{%
  \Hy@raisedlink{\org@hypertarget{#1}{}}#2%
  }
\makeatother
\usepackage[figure,table]{hypcap}
\usepackage{MnSymbol}
\usepackage{enumerate}
\usepackage{float}
\usepackage{comment}

\usepackage{subfigure}

\hypersetup{
	bookmarksnumbered,
	pdfstartview={FitH},
	citecolor={darkgreen},
	linkcolor={darkred},
	urlcolor={darkblue},
	pdfpagemode={UseOutlines}}
\definecolor{darkgreen}{RGB}{50,190,50}
\definecolor{darkblue}{RGB}{0,0,190}
\definecolor{darkred}{RGB}{238,0,0}
\usepackage{soul}

\newcommand{\ket}[1]{\ensuremath{\left|\right.\!{#1}\!\left.\right\rangle}}

\newcommand{\bra}[1]{\ensuremath{\left\langle\right.\!{#1}\!\left.\right|}}

\newcommand{\scpr}[2]{\ensuremath{\left\langle\right.\hspace*{-1pt} #1 \hspace*{-1pt}\left|\right.\hspace*{-1pt} #2 \hspace*{-1pt}\left.\right\rangle}}

\newcommand{\nl}{\ensuremath{\hspace*{-0.5pt}}}
\newcommand{\nr}{\ensuremath{\hspace*{0.5pt}}}

\newcommand{\suptiny}[3]{\ensuremath{^{\hspace{#1 pt}\protect\raisebox{#2 pt}{\tiny{$ #3$}}}}}

\newcommand{\expval}[1]{\ensuremath{\left\langle\right.\hspace*{-1pt} #1 \hspace*{-1pt}\left.\right\rangle}}

\newcommand{\tr}{\textnormal{Tr}}
\newcommand{\djj}{d\kern-0.4em\char"16\kern-0.1em}

\newtheorem{prop}{Proposition}

\renewcommand{\thesection}{\Roman{section}}
\renewcommand{\thesubsection}{\Roman{section}.\Alph{subsection}}
\renewcommand{\thesubsubsection}{\Roman{section}.\Alph{subsection}.\arabic{subsubsection}}
\makeatletter
\renewcommand{\p@subsection}{}
\renewcommand{\p@subsubsection}{}
\makeatother

\usepackage{filecontents}

\begin{document}
\title{Bayesian parameter estimation using Gaussian states and measurements}
\author{Simon Morelli}
\email{simon.morelli@oeaw.ac.at}
\affiliation{Institute for Quantum Optics and Quantum Information - IQOQI Vienna, Austrian Academy of Sciences, Boltzmanngasse 3, 1090 Vienna, Austria}
\author{Ayaka Usui}
\email{ayaka.usui@oist.jp}
\affiliation{Quantum Systems Unit, Okinawa Institute of Science and Technology Graduate University, Okinawa, Japan}
\author{Elizabeth Agudelo}
\email{elizabeth.agudelo@oeaw.ac.at}
\affiliation{Institute for Quantum Optics and Quantum Information - IQOQI Vienna, Austrian Academy of Sciences, Boltzmanngasse 3, 1090 Vienna, Austria}
\author{Nicolai Friis}
\email{nicolai.friis@univie.ac.at}
\affiliation{Institute for Quantum Optics and Quantum Information - IQOQI Vienna, Austrian Academy of Sciences, Boltzmanngasse 3, 1090 Vienna, Austria}
\date{\today}

\begin{abstract}
    Bayesian analysis is a framework for parameter estimation that applies even in uncertainty regimes where the commonly used local (frequentist) analysis based on the Cram{\'e}r-Rao bound is not well defined. 
    In particular, it applies when no initial information about the parameter value is available, e.g., when few measurements are performed. 
    Here, we consider three paradigmatic estimation schemes in continuous-variable quantum metrology (estimation of displacements, phases, and squeezing strengths) and analyse them from the Bayesian perspective.
    For each of these scenarios, we investigate the precision achievable with single-mode Gaussian states under homodyne and heterodyne detection. 
    This allows us to identify Bayesian estimation strategies that combine good performance with the potential for straightforward experimental realization in terms of Gaussian states and measurements. 
    Our results provide practical solutions for reaching uncertainties where local estimation techniques apply, thus bridging the gap to regimes where asymptotically optimal strategies can be employed. 
\end{abstract}
\maketitle
%%%%%%%%%%%%%%%%%%%%%%%%%%%%%%%%%%%%%%%%%%%%%%%%%%%%%%%%%%%%%%%%%%%%%%%%%%%%%%%%%%%%%%%

\section{Introduction}\label{sec:introduction}
\vspace*{-1mm}

    Quantum sensing devices hold the promise of outperforming their classical counterparts.  
    However, since classical strategies can achieve arbitrary precision, provided that sufficiently many independent probes are used, the advantage of quantum sensing devices does not lie in the achievable precision. 
    Instead, quantum strategies provide a faster increase in precision with $n$, the number of probes. 
    In an idealised quantum sensing scenario, the estimation precision can in principle scale at the so-called Heisenberg limit (HL) of $1/n$ as $n\rightarrow\infty$.
    In contrast, classical strategies can at most achieve a precision scaling of $1/\sqrt{n}$, the so-called standard quantum limit (SQL).
 
    In the context of quantum optics, which we are interested in here, the possibility of preparing states with uncertain photon number means that the number of probes is uncertain.
    Therefore, the scaling usually refers to resources such as the mean photon number or mean energy of the probe systems.
    Nevertheless, general quantum strategies can result in a quadratic scaling advantage and thus outperform `classical' strategies using the same resources.
    However,  two important factors have to be considered.
    
    First, preparing optimal or at least close to optimal probes and carrying out the corresponding joint measurements can be complicated and technologically demanding.
    Moreover, in the presence of uncorrelated noise the scaling advantage with increasing $n$ persists only up to a certain point, beyond which only a (potentially high) constant advantage remains~\cite{DemkowiczDobrzanskiKolodynskiGuta2012, EscherDavidovichZaguryDeMatosFilho2012, ChavesBraskMarkiewiczKolodynskiAcin2013}.
    Even if one disregards any additional costs that might incur from trying to combat noise~\cite{SekatskiSkotiniotisDuer2016, SekatskiSkotiniotisKolodynskiDuer2017}, overheads from complex preparation procedures and the resulting low probe state fidelities may thus invalidate the expected benefits.
    Consequently, it is important to identify estimation strategies that can outperform `classical' approaches while being feasibly implementable as well as robust against noise.
    For instance, for estimation problems in continuous-variable (CV) systems, Gaussian states and measurements are generally considered to be comparably easily implementable.
    They allow achieving the HL for many scenarios within the local, also called `frequentist', paradigm, including the local estimation of phases, displacements, squeezing and others~\cite{GaibaParis2009, PinelJianTrepsFabreBraun2013, Monras2013, Jiang2014, FriisSkotiniotisFuentesDuer2015, SafranekLeeFuentes2015, SafranekFuentes2016, RigovaccaEtAl2017, Safranek2019, Oh2019b}.
    
    Second, many of these insights are based on the Cram{\'e}r-Rao bound (CRB). 
    The CRB applies for estimation with unbiased estimators.
    It provides a lower bound for the precision via the inverse Fisher information (FI). 
    Estimators that are unbiased locally (i.e., for specific parameter values) are readily available, but profiting from their unbiasedness requires precise prior information on the estimated parameter. 
    The `local' approach is therefore only well-justified when the number of independent probes is sufficiently large (hence `frequentist'), in such a case, the CRB provides the asymptotically achievable limit on scaling. 
    However, when the available number of probes is limited (some authors~\cite{RubioKnottDunningham2018, RubioDunningham2019, RubioDunningham2020} refer to 'limited data' in this context) then local estimation is not well defined. 
    Resulting pathologies can lead to scaling seemingly better than the HL~\cite{RivasLuis2012, Berry2012} and even to an unbounded FI for finite average photon numbers~\cite{ZhangJinCaoLiuFan2013}. 
    The available prior information also has to be carefully considered when calculating the CRB. 
    For instance, for phase estimation with $N00N$-states, a growing (average) photon number $n$ implicitly assumes that the prior interval is narrowing with $2\pi/n$. 
    If this is not accounted for, part of the scaling advantage comes from the increasing prior information, as pointed out in Refs.~\cite{JarzynaDemkowiczDobrzanski2015, GoreckiDemkowiczDobrzanskiWisemanBerry2020}. 

    This motivates the study of Bayesian estimation approaches for quantum sensing, which we consider here. 
    In Bayesian estimation, one's initial knowledge of the parameter is described by a probability distribution (the prior) which is updated as more measurement data becomes available. 
    The Bayesian approach is valid for an arbitrary number of probes and can in this sense be considered to be more rigorous than local estimation, at the cost of introducing a dependence on the prior.
    However, the influence of the prior vanishes for larger number of measurements, since the prior knowledge becomes less and less relevant with growing amount of measurement data. 
    In practice, one may pursue a hybrid strategy, where initial Bayesian estimation is employed to sufficiently narrow down the possible range of the parameter before switching to a local estimation strategy with many repetitions.

    Here, we consider Bayesian estimation scenarios for quantum optical fields.
    While much progress has been made for CV parameter estimation within the local paradigm, in particular, regarding the calculation of the quantum Fisher information (QFI)~\cite{GaibaParis2009, PinelJianTrepsFabreBraun2013, Monras2013, Jiang2014, SafranekLeeFuentes2015, SafranekFuentes2016, RigovaccaEtAl2017, Safranek2019, Oh2019b} and the associated optimal strategies achieving the CRB~\cite{Paris2009, AspachsCalsamigliaMunozTapiaBagan2009, TothApellaniz2014, DemkowiczDobrzanskiJarzynaKolodynski2015, PezzeSmerziOberthalerSchmiedTreutlein2018, SidhuKok2020}, CV parameter estimation in the Bayesian setting is much less explored.
    There, recent work has provided insight into Bayesian estimation with discrete~\cite{FidererSchuffBraun2020} and CV systems using some specific probe states, including coherent states~\cite{RubioKnottDunningham2018, RubioDunningham2019, RubioDunningham2020, Martinez-Vargas2017}, $N00N$ states~\cite{RubioKnottDunningham2018, CiminiGenoniGiananiSpagnoloSciarrinoBarbieri2020}, and single-photon states~\cite{Valeri2020}. 
    Determining efficient and practically realizable strategies for Bayesian estimation in quantum optical systems can thus be considered an important link in the development of quantum sensing technologies, which this paper aims to establish.

    Within the Bayesian paradigm, the additional freedom represented by the choice of the prior exacerbates the difficulty of determining optimal estimation strategies, making it all the more necessary to identify practically realizable strategies that can also be easily adapted.
    Here, in particular, we are interested in identifying strategies for Bayesian estimation considering Gaussian states and Gaussian measurements.
    Gaussian states not only permit an elegant mathematical description in phase space, but are also especially easy to realise experimentally and are by now broadly used~\cite{KlauderSkagerstam1985,AndersenGehringMarquardtLeuchs2016}. 
    Gaussian measurements, i.e., homodyne or heterodyne detection, have been shown to outperform number detection for few repetitions~\cite{RubioDunningham2019} and to be more robust against noise~\cite{GenoniOlivaresParis2011,GiovannettiLloydMaccone2006, DemkowiczDobrzanskiJarzynaKolodynski2015} than photon number detection or `on/off' detection---which discriminates only between the absence or presence of photons. 

    To broadly investigate the performance of Gaussian states and measurements in Bayesian metrology, we consider three paradigmatic problems: the estimation of phase-space displacements, phase estimation, and the estimation of single-mode squeezing. 
    For each task, we provide practically realisable strategies based on single-mode Gaussian states combined with homodyne or heterodyne detection that allow efficiently narrowing the prior to the point where local estimation strategies may take over. 
    To set the stage for this investigation, we briefly review the method of Bayesian estimation and relevant concepts of Gaussian quantum optics in Sec.~\ref{sec:framework}. 
    In Sec.~\ref{sec:displacent}, we focus on the estimation of displacements for Gaussian priors, and provide analytical results for the achievable precision using single-mode Gaussian states for both homodyne and heterodyne detection. 
    In Secs.~\ref{sec:phase} and~\ref{sec:squeezing}, we proceed with similar investigations of Bayesian estimation of phases and squeezing parameters, where we compare the performance of squeezing and displacement of the probe system. 
    Finally, we discuss our results and provide an outlook and conclusions in Sec.~\ref{sec:conclusions}.
%%%%%%%%%%%%%%%%%%%%%%%%%%%%%%%%%%%%%%%%%%%%%%%%%%%%%%%%%%%%%%%%%%%%%%%%%%%%%%%%%%%%%%%

\vspace*{-1mm}
\section{Framework}\label{sec:framework}
\vspace*{-1mm}

    In this section, we provide a brief overview of the relevant concepts in Bayesian estimation (Sec.~\ref{sec:Bayesian quantum estimation}) and Gaussian quantum optics (Sec.~\ref{sec:Gaussian quantum optics}), before we present our results in the following sections.
    For a more extensive overview of classical Bayesian estimation theory we refer to~\cite{DAgostini2003,Bolstad2009,AmaralTurkmanPaulinoMuller2019}, while more details on local and Bayesian estimation in the quantum setting can be found, e.g., in the appendix of~\cite{FriisOrsucciSkotiniotisSekatskiDunjkoBriegelDuer2017}.
%%%%%%%%%%%%%%%%%%%%%%%%%%%%%%%%%%%%%%%%%%%%%%%%%%%%%%%%%%%%%%%%%%%%%%%

\vspace*{-1mm}
\subsection{Bayesian quantum parameter estimation}\label{sec:Bayesian quantum estimation}
\vspace*{-1mm}

\subsubsection{The Bayesian estimation scenario}\label{sec:Bayesian estimation}
\begin{figure}[t]
        \begin{center}
            \includegraphics[width=0.95\columnwidth]{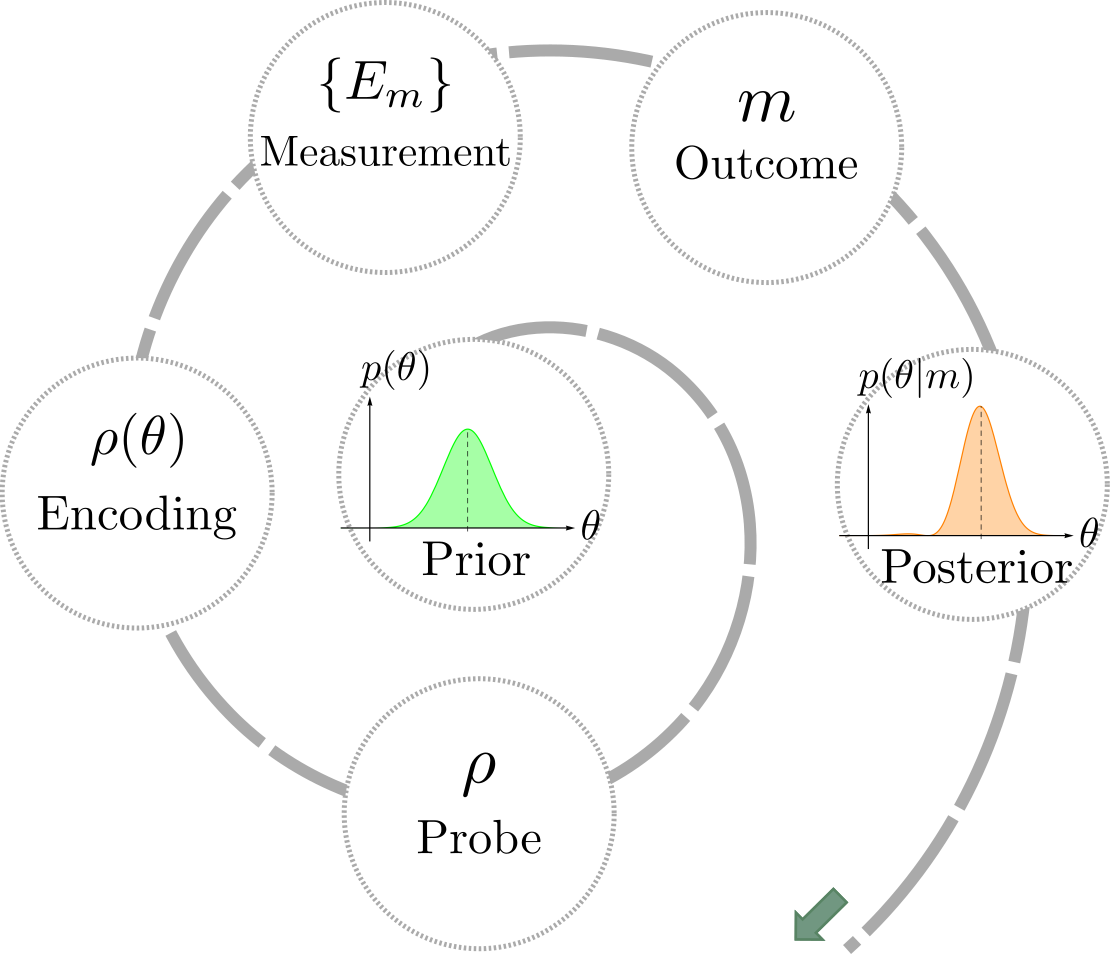}
            \caption{
            Bayesian Quantum Parameter Estimation. 
            In Bayesian estimation scenarios, prior information encoded in a probability distribution $p(\theta)$ is updated based on available measurement data such as observing a particular measurement outcome $m$, resulting in a posterior conditional probability distribution $p(\theta|m)$. 
            In quantum parameter estimation, the measurement procedure consists of preparing the system in a probe state $\rho$ on which the parameter $\theta$ is encoded by a suitable transformation. 
            The measurement is represented by a positive-operator valued measure (POVM) with elements $E_{m}$ representing the possible outcomes $m$.
            }
            \label{fig:Bayesian_estimation}
        \end{center}
    \end{figure}

\vspace*{-1mm}
    The framework of Bayesian parameter estimation revolves around updating initially available information (or a previously held belief) based on new measurement data via Bayes' theorem, as we will explain in the following.
    The initial knowledge of the estimated parameter $\theta$ is encoded in a probability distribution $p(\theta)$ called the \emph{prior distribution function} or `prior' for short. 
    It captures all our beliefs (system properties, expertise) and information (prior experimental data) about the system under investigation. 
    When a measurement is performed on the system, the probability $p(m|\nr\theta)$ to observe the measurement outcome $m$ in a system characterised by the parameter $\theta$ is called the \emph{likelihood}, and can be calculated from the properties of the model used to describe the system and the measurement.
    Combined with the prior $p(\theta)$, the likelihood leads one to expect the outcome $m$ with probability
    \begin{align}
        p(m) &=\,\int\!\! d\theta\ p(m|\nr\theta)\,p(\theta),
    \end{align}
    where the integral is over the support of the prior and it is to be understood as a sum in case of a discrete parameter. 
    The conditional probability that the estimated parameter equals $\theta$, given that measurement outcome $m$ was observed, can then be calculated via Bayes' law, i.e.,
    \begin{align}
        p(\theta|m) &=\,\frac{p(m|\nr\theta)\,p(\theta)}{p(m)}.\label{bayeslaw}
    \end{align}
    The function $p(\theta|m)$ is called the \emph{posterior distribution} of the system parameter, after we have updated our belief with newly available data.
    The updating procedure, illustrated in Fig.~\ref{fig:Bayesian_estimation}, can be repeated arbitrary many times, where the posterior of one step serves as the prior in the next step and the measurement procedure leading to $p(m|\nr\theta)$ can in principle also be adapted from step to step. 

    After concluding the measurements, the posterior distribution represents a complete description of all available information about the parameter.
    Nevertheless, it is often desirable (even if not strictly necessary) to nominate an estimator $\hat{\theta}$ and a suitable variance to express the result of the estimation procedure. 
    While the estimator assigns a specific value for $\theta$ to any prior or posterior, the variance quantifies the associated uncertainty in the estimate. 
    For parameters $\theta\in\mathbb{R}$, the canonical choice for an estimator is the mean value of the posterior distribution
    \begin{align}
        \hat{\theta}(m)=\langle\theta\rangle=\int\!\! d\theta\ p(\theta|m)\,\theta.
    \end{align}
    In this case, a valid figure of merit for the confidence in this estimate is the variance of the posterior
    \begin{align}
        V_{\mathrm{post}}(m) &=\,\int\!\!d\theta\ p(\theta|m)\,\bigl[\theta-\hat{\theta}(m)\bigr]^{2}.
        \label{eq:posterior variance MSE}
    \end{align}
    A wide posterior with large variance suggests there is still high uncertainty in our belief about the parameter, whereas a narrow distribution with small variance indicates high confidence in our estimator. 
    Since the variance of the posterior generally depends on the measurement outcome, a good figure of merit for the expected confidence in the estimate provided by a particular measurement strategy is the average variance of the posterior,
    \begin{align}
        \bar{V}_{\mathrm{post}}=\int\!\!dm\ p(m)\,V_{\mathrm{post}}(m),
    \end{align}
    which we will use here to quantify the precision of the estimation process.
    However, note that in some cases, the mean and mean square error variance above need to be replaced by more appropriate quantifiers. 
    For instance, in the case that the parameter in question is a phase, where $\theta=-\pi$ and $\theta=\pi$ are identified, $\hat{\theta}(m)$ and $V_{\mathrm{post}}(m)$ can be replaced by suitable alternatives, as we will discuss in in Sec.~\ref{sec:phase}. 
    In any given setting, the task is then to determine estimation strategies that provide sufficiently high precision.

    The precision of the estimation procedure generally depends on the shape of the prior, which can in principle be an arbitrarily complicated distribution. Uninformative priors generally influence the outcome less than narrow priors, so one should always be careful which amount of information should be encoded in the prior. 
    However, the influence of the prior on the final estimate generally reduces with increasing number of measurements, and can be argued to become irrelevant asymptotically, see, e.g.,~\cite[Chapter~13]{DAgostini2003}. 
    Consequently, encoding one's knowledge only approximately using a family of probability distributions with only few degrees of freedom can help to facilitate a more straightforward evaluation of the performance of the chosen strategy, while preserving its qualitative features. 
    
    For instance, a class of probability distributions is said to be conjugate to a given likelihood function, if priors from within this class result in posterior distributions that belong to that class as well.
    Choosing the prior to be conjugate to the likelihood in this way makes the updating particularly easy, since this only requires the parameters to be updated to define the posterior distribution uniquely within the chosen class of probability functions, instead of requiring an entirely new calculation to determine the posterior. 
    Gaussian distributions are self-conjugate with respect to the mean, e.g. for Gaussian likelihood functions encoding the parameter to be estimated in their mean, the class of conjugate priors are Gaussian distributions as well.
    The following proposition is a well known result in statistical theory~\cite{RaiffaSchlaifer1961,DAgostini2003,Bolstad2009,AmaralTurkmanPaulinoMuller2019}.

    \begin{prop}\label{conjugatedprior}
        Let the likelihood be Gaussian distributed, $p(m|\nr\theta)=\mathcal{N}_{m}\bigl(\bar{m}(\theta),\tilde{\sigma}^2\bigr)\propto\mathcal{N}_{\theta}\bigl(\bar{\theta}(m),\sigma^2\bigr)$, where $\bar{\theta}(m)$ is the mean of the distribution in $\theta$, the parameter to be estimated. 
        Then a Gaussian prior is the natural conjugate, i.e., if the prior is Gaussian distributed with $p(\theta)=\mathcal{N}_{\theta}(\mu_0,\sigma_0^2)$, the posterior distribution $p(\theta|m)$ is also Gaussian with mean value $\mu_p=\bigl[\sigma^2\mu_0 + \sigma_0^2\bar{\theta}(m)\bigr]/(\sigma_0^2+\sigma^2)$ and variance $\sigma_p^2=(\sigma^2\sigma_0^2)/(\sigma_0^2+\sigma^2)$.
    \end{prop}

%%%%%%%%%%%%%%%%%%%%%%%%%%%%%%%%%%%%%%%%%%%%%%%%%%%%%%%%%%%%%%%

\vspace*{-1mm}
\subsubsection{Bayesian estimation using quantum systems}
\label{sec:quantum estimatio in Bayesian setting}
\vspace*{-1mm}
    The framework of Bayesian estimation can easily be applied to a quantum setting, as illustrated in Fig.~\ref{fig:Bayesian_estimation}. 
    In this case the parameter $\theta$ one is interested in estimating is encoded by a transformation that can generally be a completely positive and trace-preserving (CPTP) map. 
    However, in many cases, including those we study here, the transformation is considered to be a unitary $U_\theta$ that acts on an initially prepared probe state, represented by a density operator $\rho$. 
    The resulting encoded state is then given by $\rho(\theta)=U_{\theta}\rho U^{\dagger}_{\theta}$. 
    The measurement of the encoded state can then be represented by a positive operator-valued measure (POVM) with elements $E_{m}\geq0$, whose integral (or sum in case of a discrete set of possible measurement outcomes $m$) evaluates to the identity on the Hilbert space of the probe, i.e., $\int\!\! dm\,E_{m}=\mathds{1}$. 
    In the quantum case the likelihood is then given by
    $ p(m|\nr\theta) =\,\tr\bigl[E_{m}\rho(\theta)\bigr]$.

    In local estimation scenarios with unbiased estimators $\hat{\theta}$, the CRB gives a lower bound for the variance of the estimator in terms of the inverse Fisher information $I\bigl[p(m|\nr\theta)\bigr]$, that is, 
    $V(\hat{\theta})\ge I[p(m|\nr\theta)]^{-1}$.
    Here, the Fisher information depends only on the likelihood function and is given by
    \begin{align}
        I\bigl[p(m|\nr\theta)\bigr] &=\,
        \int\!\!dm\,p(m|\nr\theta)\,
        \Bigl[\tfrac{\partial}{\partial\theta}\log p(m|\nr\theta)\Bigr]^{2}.
    \end{align}
    In the asymptotic limit of infinite sample size, the CRB is always tight, since it is saturated by the maximum likelihood estimator, which becomes unbiased in this limit, see e.g.,~\cite{Frieden2004}. 
    Any local estimation problem can thus be reduced to determining an estimation strategy with a likelihood $p(m|\nr\theta)$ corresponding to as large a FI as possible. 
    In the quantum setting, this leaves us with the task of determining suitable probe states $\rho$ and measurements $\{E_{m}\}_{m}$. 
    The optimisation of the FI over all POVMs can be carried out analytically, leading to the quantum Fisher information (QFI) $\mathcal{I}\bigl[\rho(\theta)\bigr]$,
    and the corresponding quantum CRB~\cite{DemkowiczDobrzanskiJarzynaKolodynski2015, BraunsteinCaves1994}, $V(\hat{\theta})\geq 1/\mathcal{I}\bigl[\rho(\theta)\bigr]$. 
    The QFI can be expressed in terms of the Uhlmann fidelity $\mathcal{F}(\rho_{1},\rho_{2})=\bigl(\tr\sqrt{\sqrt{\rho_{1}}\rho_{2}\sqrt{\rho_{1}}}\bigr)^{2}$ as
    \begin{align}
        \mathcal{I}\bigl[\rho(\theta)\bigr]  &=\,\lim_{d\theta\rightarrow0}8\,\frac{1-\sqrt{\mathcal{F}\bigl[\rho(\theta),\rho(\theta+d\theta)\bigr]}}{d\theta^{2}}.
    \end{align}
    
    For the Bayesian estimation scenario, a similar bound exists. 
    The \textit{Van Trees inequality} bounds the average variance from below according to
    \begin{equation} \label{eq:vantrees}
        \bar{V}_{\mathrm{post}}\ge \frac{1}{I\bigl[p(\theta)\bigr]+\bar{I}\bigl[p(m|\nr\theta)\bigr]},
    \end{equation}
    where $I\bigl[p(\theta)\bigr]=\int\!\!d\theta\,p(\theta)\,\bigl[\tfrac{\partial}{\partial\theta}\log p(\theta)\bigr]^{2}$ is the FI of the prior and $\bar{I}\bigl[p(m|\theta)\bigr]=\int\!\! d\theta\ I\bigl[p(m|\nr\theta)\bigr]\,p(\theta)$ is the average FI of the likelihood~\cite{VanTrees2001, Kay1993}. 
    This inequality is often referred to as the Bayesian Cram{\'e}r-Rao bound, see, e.g.,~\cite{GillLevit1995}. 
    In contrast to the CRB in the local scenario, this bound is not tight, which means there might not exist a strategy achieving the equality.

    In a Bayesian quantum estimation problem, the Van Trees inequality can be modified to a Bayesian version of the quantum CRB by noting that the FI is bounded from above by the QFI, $\mathcal{I}\bigl[\rho(\theta)\bigr]\geq I\bigl[p(m|\nr\theta)\bigr]$. 
    Moreover, if the parameter to be estimated is encoded by a unitary transformation $U_\theta$, the QFI is independent of $\theta$. 
    Consequently, the average FI can be bounded by the QFI to obtain the Bayesian quantum CRB
    \begin{equation} \label{eq:QBCR_bound}
        \bar{V}_{\mathrm{post}}\ge \frac{1}{I\bigl[p(\theta)\bigr]+\mathcal{I}\bigl[\rho(\theta)\bigr]},
    \end{equation}
    which gives a lower bound for the average variance for all possible POVMs~\cite{FriisOrsucciSkotiniotisSekatskiDunjkoBriegelDuer2017}. 
    As before with Eq.~(\ref{eq:vantrees}), this bound is not tight. 

    While well-known methods for constructing optimal POVMs for fixed probe states exist for local estimation, optimization of the probe state and measurements for Bayesian estimation has to be carried out on a case-by-case basis and is typically challenging. 
    At the same time, states and measurements that are optimal for a given prior may require complicated preparation procedures while generally no longer being optimal after even a single update. 
    Consequently, it is of interest to devise measurement strategies for Bayesian estimation that are easily realizable and provide `good' performance for different priors. 
    Here, we provide and examine such strategies for a range of estimation problems in quantum optical scenarios.
%%%%%%%%%%%%%%%%%%%%%%%%%%%%%%%%%%%%%%%%%%%%%%%%%%%%%%%%%%%%%%%%%%%%%%%

\subsection{Gaussian quantum optics}\label{sec:Gaussian quantum optics}

    As we established before, we are interested in the analysis of scenarios where probe states are quantum states of the electromagnetic field. 
    In particular, our goal is studying the performance of Gaussian states. 
    To set the stage for this investigation, we will here briefly summarize the relevant concepts of Gaussian quantum optics. 
    For a more extensive treatment of CV systems and Gaussian quantum optics we refer the reader to the Refs.~\cite{VogelWelsch2006,GerryKnight2005} and for the particular context of quantum information processing cf. Refs.~\cite{BraunsteinVanLoock2005, FerraroOlivaresParis2005, AdessoIlluminati2007, WangHiroshimaTomitaHayashi2007, WeedbrookPirandolaGarciaPatronCerfRalphShapiroLloyd2012, AdessoRagyLee2014}. 
    Multimode optical fields can be represented as collections of bosonic modes. We consider a CV system that consists of $N$ bosonic modes, i.e., $N$ quantum harmonic oscillators. 
    To each mode, labelled $k$, one associates a pair of annihilation and creation operators, $\hat{a}_k$ and $\hat{a}_k^\dagger$, respectively. These mode operators satisfy the bosonic commutation relations $[\hat{a}_k, \hat{a}^\dagger_l]= \delta_{kl}$.
    The mode operators can be combined into the quadrature operators $\hat{q}_k=(\hat{a}_k+\hat{a}_k^\dagger)/\sqrt{2}$ and $\hat{p}_k=i(\hat{a}_k^\dagger-\hat{a}_k)/\sqrt{2}$. 
    These operators correspond to the generalized position and momentum observables for the mode $k$. 
    They have continuous spectra, and eigenbases $\{|q\rangle\}_{q\in \mathbb{R}}$ and $\{|p\rangle\}_{p\in \mathbb{R}}$, respectively. 
    In the simplectic form~\cite{Arvind1995}, the quadrature operators are collected in one single vector $\mathbf{\hat{x}}=(\hat{q}_1,\hat{p}_1,\ldots,\hat{q}_N,\hat{p}_N)^T$. 
    
    The state of such an $N$-mode system is described by a density operator $\rho \in \mathcal{D}(\mathcal{H}^{\otimes N})$, a positive (semi-definite) and unit trace operator. 
    Alternatively, the state of the system can be represented by its Wigner function $W(\mathbf{x})$ \cite{Wigner1932}, i.e., a quasiprobability distribution in the $2N$-dimensional phase space with real coordinates $q_{i},p_{i}\in\mathbb{R}$, collected in a vector $\mathbf{x}=(q_{1},p_{1},\ldots,q_{N},p_{N})^{T}$. 
    
\vspace*{-1mm}
\subsubsection{Gaussian states}
\label{sec:gaussian_states}
\vspace*{-1mm}
    In the cases where the Wigner function of the state is a multivariate Gaussian distribution of the form
    \begin{align} \label{eq:wignerfcn}
        W(\mathbf{x})=\frac{\exp[-(\mathbf{x}-\mathbf{\bar{x}})^T\mathbf{\Gamma}^{-1}(\mathbf{x}-\mathbf{\bar{x}})]}{\pi^N\sqrt{\det(\mathbf{\Gamma})}},
    \end{align}
    the states are called \textit{Gaussian}.
    Gaussian states are fully characterized by its vector of first moments $\mathbf{\bar{x}}=\tr(\mathbf{\hat{x}}\rho)$ and its covariance matrix $\mathbf{\sigma}=(\sigma_{ij})=\tfrac{1}{2}\mathbf{\Gamma}$.
    The real and symmetric $2N\times2N$ covariance matrix collects the second moments $\sigma_{ij}=\langle\{\mathbf{\hat{x}}_i-\langle\mathbf{\hat{x}}_i\rangle,\mathbf{\hat{x}}_j-\langle\mathbf{\hat{x}}_j\rangle\}\rangle/2$. 
    Examples for Gaussian states include the vacuum state, thermal states as used, e.g., to describe black-body radiation, or coherent states modelling the photon distribution in a laser.
    The full description via the vector of first moments and the covariance matrix allows one to completely and compactly capture an important class of familiar states in an infinite-dimensional Hilbert space via a finite number of degrees of freedom.

    In this paper we investigate the performance of single-mode Gaussian states for Bayesian parameter estimation. 
    More specifically, we consider coherent and displaced-squeezed states. 
    Coherent states are the right-eigenstates of the annihilation operator $\hat a_k$ such that $\hat a_k\ket{\alpha}_k=\alpha\ket{\alpha}_k$ and form a basis in the Hilbert space $\mathcal{H}_k$. 
    They result from applying the displacement operator of the coherent amplitude $\alpha\in\mathbb{C}$, 
    \begin{equation}
        \hat D_k(\alpha) = \exp\big(\alpha\hat a^\dagger_k-\alpha^*\hat a_k\big),
        \label{eq:displacement_op}
    \end{equation}
    to the vacuum $\ket{0}_k$, such that $\ket{\alpha}_k=\hat D_k(\alpha)\ket{0}_k$. 
    Coherent states are states with the same covariance matrix as the vacuum state. 
    For a single-mode coherent state $\ket{\alpha}_k$, the first moment is $\mathbf{\bar{x}}=\sqrt{2}\bigl[\Re(\alpha),\Im(\alpha)\bigr]^{T}$ and the second moment is the identity matrix divided by 2, meaning that the variance both in $\hat{q}_k$ and $\hat{p}_k$ equals $1/2$, saturating the uncertainty relation in a balanced way. 
    
\begin{figure}[t]
        \begin{center}        \includegraphics[width=0.95\columnwidth]{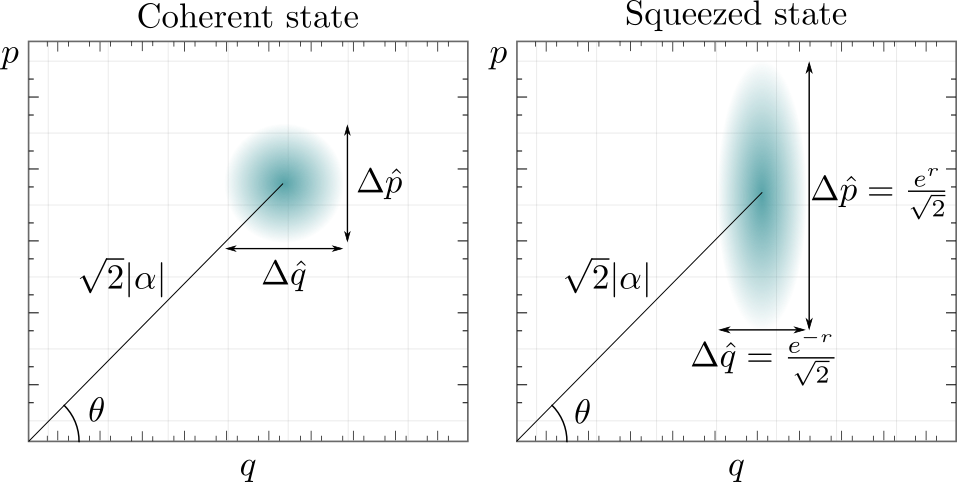}
            \caption{
            Contours of the Wigner functions for single-mode Gaussian states. 
            The Wigner functions are given by Gaussian distributions of the form Eq.~(\ref{eq:wignerfcn}), and are characterised by a complex displacement $\alpha$, a real squeezing strength $r$ and a squeezing angle $\varphi$. The illustration compares a displaced vacuum state ($r=0$) on the left-hand side and a squeezed displaced state with $r>0$ and $\varphi=0$ on the right-hand side. The width of the latter Wigner function is reduced in the $\hat{q}$-quadrature and increased in the $\hat{p}$-quadrature with respect to the coherent state.
            }
        \end{center}
        \label{fig:wigner_gaussian}
    \end{figure}

    Coherent states are not the only states saturating the uncertainty relation. 
    Indeed, squeezed states are a larger class of states with this property, while allowing for unbalanced variances of the two canonical quadratures for each mode, c.f. Fig. \ref{fig:wigner_gaussian}.
    Squeezed states are obtained by the action of the squeezing operator, 
    \begin{equation}
        \hat S_k(\xi)=\exp\left[\frac{1}{2}(\xi^*\hat{a}_k^2-\xi\hat{a}_k^{\dagger 2})\right],
        \label{squeezingoperator}
    \end{equation}
    on the vacuum $\ket{0}_k$. 
    The states $\hat S_k(\xi)\ket{0}_k$ are characterized by a complex parameter $\xi=r e^{i\varphi}$, where $r\in\mathbb{R}$ is the so-called squeezing strength, and $\varphi\in [0,2\pi)$ is the squeezing angle. 
    
    Every pure single-mode Gaussian state has minimal uncertainty and can be generated by the combined action of squeezing and displacement operators on the vacuum state. 
    Such states are therefore entirely specified by their displacement parameter $\alpha\in\mathbb{C}$, their squeezing strength $r\in\mathbb{R}$, and their squeezing angle $\varphi\in [0,2\pi)$.
    If squeezing is restricted to a real parameter only, then also a phase rotation
    \begin{equation}
        \hat{R}_k(\theta)=\exp\big(-i\theta\hat{a}^\dagger_k\hat{a}_k\big),
        \label{eq:phase rotation}
    \end{equation}
    is needed to describe the most general pure single-mode Gaussian state. 
    The vector of first moments of such a displaced squeezed state $|\alpha, r \mathrm{e}^{i\varphi}\rangle=\hat{D}(\alpha) \hat{S}(\xi) |0\rangle = \hat{D}(\alpha) \hat{R}(\varphi/2) \hat{S}(r)|0\rangle$ 
    is given by $\mathbf{\bar{x}}=\sqrt{2}\bigl[\Re(\alpha),\Im(\alpha)\bigr]^{T}$ and its covariance matrix is
    \begin{align}
        \mathbf{\sigma}=\frac{1}{2}\begin{pmatrix} \cosh{2r} - \cos{\varphi}\ \sinh{2r} &\hspace*{4mm}\sin{\varphi}\ \sinh{2r}\\ 
        \hspace*{-12mm}\sin{\varphi}\ \sinh{2r} &\hspace*{-8mm}\cosh{2r} + \cos{\varphi}\ \sinh{2r} \end{pmatrix}.
        \label{eq:cov matrix single mode states}
    \end{align}

    A unitary transformation is called Gaussian, if it maps Gaussian states into Gaussian states. 
    This class of unitary operations is generated by
    Hamiltonians that are (at most) second order polynomials of the mode operators. 
    Notice that every single-mode Gaussian unitary operation can be decomposed into displacement, rotation, and squeezing operations. 
    In addition to having a relatively straightforward theoretical description, Gaussian states and Gaussian transformations are also especially relevant in practice, since they are typically easy to produce and manipulate experimentally~\cite{KlauderSkagerstam1985,AndersenGehringMarquardtLeuchs2016}.
    
\vspace*{-1mm}
\subsubsection{Gaussian measurements}
\label{sec:gaussian_measurements}
\vspace*{-1mm}
    Any measurement can be described by a positive-operator valued measure (POVM). 
    In CV quantum information, it is common to use continuous POVMs, that is, POVMs that are continuous sets of operators and a continuous range of measurement outcomes. 
    A measurement is called Gaussian if it gives a Gaussian distribution of outcomes whenever it is applied to a Gaussian state. 
    Gaussian measurements that are frequently considered in the context of CV quantum information are homodyne \cite{YuenShapiro1978, ShapiroYuenMata1979} and heterodyne detection \cite{YuenShapiro1980}.
    Homodyne detection corresponds to the measurement of a mode quadrature, for example $\hat{q}$. 
    In this case, the POVM consists of projectors onto the quadrature basis,  $\{|q\rangle\!\langle q|\}_{q\in\mathbb{R}}$. 
    For heterodyne detection the POVM elements are projectors onto coherent states $\{\frac{1}{\pi}|\beta\rangle\!\langle \beta|\}_{\beta\in\mathbb{C}}$.
    Moreover, we note that it has recently been shown that every bosonic Gaussian observable can be considered as a combination of (noiseless and noisy) homodyne and heterodyne detection~\cite{Holevo2020}.
%%%%%%%%%%%%%%%%%%%%%%%%%%%%%%%%%%%%%%%%%%%%%%%%%%%%%%%%%%%%%%%%%%%%%%%%%%%%%%%%%%%%%%%

\section{Displacement estimation}
\label{sec:displacent}
    We now consider Bayesian estimation of displacements using Gaussian states and Gaussian measurements. 
    That is, we assume a displacement operator $\hat{D}(\alpha)$ as in Eq.~(\ref{eq:displacement_op}) acts on our system, initially prepared in a Gaussian probe state. 
    We then want to estimate the unknown displacement parameter $\alpha=\alpha_{\mathrm{R}}+i\alpha_{\mathrm{I}}$, with $\alpha_{\mathrm{R}}, \alpha_{\mathrm{I}}\in\mathbb{R}$. 
    To this end, we focus on estimation strategies based on heterodyne and homodyne detection. 
    These measurements are covariant under the action of displacement in the sense that the probability distribution obtained by displacing the probe state gives the same probability distribution translated by the displacement parameter in the parameter space~\cite{ChiribellaDarianoPerinottiSacchi2004}. 
    Without loss of generality, we can therefore assume that the initial probe state has not been displaced from the origin, i.e., that our probe state is a squeezed vacuum state $|\xi\rangle=\hat{S}(\xi)|0\rangle$ with $\hat{S}(\xi)$ defined in Eq.~(\ref{squeezingoperator}). 
    We further assume that our prior knowledge of the displacement is encoded in a Gaussian distribution of width $\sigma_{0}$ that is centered around $\alpha_{0}$, i.e.,
    \begin{align}
        p(\alpha)=\frac{1}{2 \pi  \sigma_0^2}\exp\left(-\frac{|\alpha-\alpha_{0}|^2}{2 \sigma_0^2}\right).\label{priordisplacement}
    \end{align}
    Our goal is then to examine the performance of the estimation strategies based on heterodyne and homodyne detection, including the respective asymptotic behaviour, both in the limit of high photon numbers and of repeated measurements, and compare the respective results. 
%%%%%%%%%%%%%%%%%%%%%%%%%%%%%%%%%%%%%%%%%%%%%%%%%%%%%%%%%

\subsection{Heterodyne measurement}
\label{eq:displacement estimation heterodyne}
    Let us first consider heterodyne detection, where the measurement is described by the POVM $\{\frac{1}{\pi}|\beta\rangle\!\langle\beta|\}_{\beta\in\mathbb{C}}$.
    The probability to obtain the measurement outcome $\beta$, given a displacement of $\alpha$, is 
    \begin{align}
        p(\beta\nr|\nr\alpha) &=\tfrac{1}{\pi}\tr\left[|\beta\rangle\!\langle\beta|\hat{D}(\alpha)|\xi\rangle\!\langle \xi|\hat{D}^\dagger (\alpha)\right]
        = \tfrac{1}{\pi}\mathcal{F}\bigl(|\beta-\alpha\rangle,|\xi\rangle\bigr).
        \label{likelihood_dis_het}
    \end{align}
    Here, $\mathcal{F}\bigl(\rho_{1},\rho_{2}\bigr)$ is the Uhlmann fidelity of the states $\rho_{1}$ and $\rho_{2}$ (defined in Sec. \ref{sec:quantum estimatio in Bayesian setting}), which reduces to  $\mathcal{F}\bigl(\ket{\psi},\ket{\phi}\bigr)=|\scpr{\psi}{\phi}|^{2}$ for pure states. 
    For two Gaussian states, the fidelity can be written in terms of the respective first moments $\mathbf{\bar{x}}_{1}$ and $\mathbf{\bar{x}}_{2}$, and second moments $\mathbf{\Gamma}_{1}$ and $\mathbf{\Gamma}_{2}$ (cf.~\cite{PinelJianTrepsFabreBraun2013}) as
    \begin{align}
        \mathcal{F}(\rho_1,\rho_2)  &= 
        \tfrac{2 \exp[
        -(\mathbf{\bar{x}}_{1}-\mathbf{\bar{x}}_{2})^{T} 
        (\mathbf{\Gamma}_{1}+\mathbf{\Gamma}_{2})^{-1} 
        (\mathbf{\bar{x}}_{1}-\mathbf{\bar{x}}_{2})
        ]}
        {\sqrt{|\mathbf{\Gamma}_1+\mathbf{\Gamma}_2|+(1-|\mathbf{\Gamma}_1|)(1-|\mathbf{\Gamma}_2|)}-\sqrt{(1-|\mathbf{\Gamma}_1|)(1-|\mathbf{\Gamma}_2|)}}.
        \label{eqfidelity}
    \end{align}
    
    For simplicity we now assume that our probe state is squeezed only along one fixed direction, i.e., $\varphi=0$. 
    This simplifies the following calculation considerably. In particular, this allows us to write the fidelity, the likelihood, and posterior distribution as products of the corresponding distributions for the real and imaginary part of the displacements, respectively. 
    In contrast, for the general case of probe states squeezed along arbitrary directions, the resulting formulas are unwieldy and complicated, but qualitatively yield the same behaviour as for $\varphi=0$. 
    We therefore refrain from presenting these calculations here. 
    
    In our case, we have  $\rho_{1}=\ket{\beta-\alpha}\!\!\bra{\beta-\alpha}$ and $\rho_{2}=\ket{\xi}\!\!\bra{\xi}$, for which the first moments are
    \begin{equation*}
        \mathbf{\bar{x}}_{\beta-\alpha}=\sqrt{2}\begin{pmatrix} \Re[\beta-\alpha]\\ \Im[\beta-\alpha]\end{pmatrix}=\sqrt{2}\begin{pmatrix} \beta_{\mathrm{R}}-\alpha_{\mathrm{R}}\\ \beta_{\mathrm{I}}-\alpha_{\mathrm{I}}\end{pmatrix}
        \ \  \text{and}\ \ 
        \mathbf{\bar{x}}_{\xi}=\begin{pmatrix} 0\\ 0\end{pmatrix},
    \end{equation*}
    while the second moments are represented by
    \begin{equation*}
    \mathbf{\Gamma}_{\beta-\alpha}=\mathds{1}_{2} 
    \ \ \ \  \text{and}\ \ \ \ 
    \mathbf{\Gamma}_\xi=\begin{pmatrix}\mathrm{e}^{-2r} & 0\\ 
        0 & \mathrm{e}^{2r}\end{pmatrix},
    \end{equation*}
    respectively.
    Accordingly, $p(\beta\nr|\nr\alpha)$ from Eq.~(\ref{likelihood_dis_het}) becomes
    \begin{align}
    p(\beta\nr|\nr\alpha) &=
    \frac{\exp\big[-\frac{\mathrm{e}^{r}(\beta_{\mathrm{R}}-\alpha_{\mathrm{R}})^2+\mathrm{e}^{-r}(\beta_{\mathrm{I}}-\alpha_{\mathrm{I}})^2}{\cosh{r}}\big]}{\pi\cosh{r}} \nonumber\\[1mm]
    &=\,
    p(\beta_{\mathrm{R}}|\alpha_{\mathrm{R}})\,p(\beta_{\mathrm{I}}|\alpha_{\mathrm{I}}),
    \end{align}
    where the distributions  $p(\beta_i|\alpha_i)$ for $i=$ R, I are given by 
    \begin{align}
    p(\beta_i|\alpha_i) = \frac{\sqrt{2}\exp\left[-\frac{2(\beta_i-\alpha_i)^2}{1+\mathrm{e}^{\mp2r}}\right]}{\sqrt{\pi(1+\mathrm{e}^{\mp2r})}}.
    \end{align}
    Here and in the following equations, the upper and lower signs in $\pm$ and $\mp$ correspond to the subscripts $i=\mathrm{R}$ and $i=\mathrm{I}$, respectively, i.e., for $i=\mathrm{R}$, the respective upper signs apply, while the lower signs apply for $i=\mathrm{I}$.
    With this expression for the likelihood and with the prior from Eq.~(\ref{priordisplacement}), one can use Bayes' law [Eq.~(\ref{bayeslaw})] to calculate the posterior distribution, the estimators and the (average) variance. 
    This allows one to evaluate the average variance for different estimation scenarios.
    We rely on such an approach in the next sections. 
    However, in the special case where both prior and likelihood are Gaussian, these two quantities are conjugate to each other. 
    Following  Prop.~\ref{conjugatedprior}, the posterior is therefore also Gaussian, and we can write down the mean and variance of the posterior directly by inspecting the likelihood and the prior.  
    That is, by noting that $\sigma^2=(1+\mathrm{e}^{\mp2r})/4$, $\mu_0=\alpha_{0,i}$, and $\bar{\theta}(m)=\beta_{i}$,
    Prop.~\ref{conjugatedprior} provides the mean and variance of the distributions $p(\alpha_{i}|\beta_{i})$.
    Again using subscripts $i=$ R, I to denote real and imaginary parts, respectively, the means are
    \begin{align}
        \hat{\alpha}_i(\beta_{i})=
        \frac{4\beta_{i}\sigma_0^2+\alpha_{0,i}(1+\mathrm{e}^{\mp2r})}{4\sigma_0^2+1+\mathrm{e}^{\mp2r}},
        \label{mean_dis_het}
    \end{align}
    which we choose as estimators for the real and imaginary part of the parameter $\alpha$, and the variances are
    \begin{align}
        \operatorname{Var}[p(\alpha_{i}|\beta_{i})]
        =\left[\frac{1}{\sigma_0^2}+2(1\pm\tanh{r})\right]^{-1}.
        \label{eq:var_dis_het}
    \end{align}
    We then define the total variance of the posterior $p(\alpha\nr|\nr\beta)$ for the complex parameter $\alpha$ as 
    \begin{align}
        \operatorname{Var}[p(\alpha\nr|\nr\beta)]=\int\!\!d\alpha\ 
        p(\alpha\nr|\nr\beta)\,|\alpha-\hat{\alpha}(\beta)|^{2}.
        \label{var2_dis_het_definition}
    \end{align}
    Because the real and imaginary parts become independent, we can further write
    the total variance as the sum of the variances of the two independent estimation parameters, i.e.,
    \begin{align}
        \operatorname{Var}[p(\alpha\nr|\nr\beta)]=\operatorname{Var}[p(\alpha_{\mathrm{R}} |\beta_{\mathrm{R}})]+\operatorname{Var}[p(\alpha_{\mathrm{I}}|\beta_{\mathrm{I}})].
        \label{var2_dis_het}
    \end{align}
    After inserting Eq.~(\ref{eq:var_dis_het}) twice, the latter expression is independent of $\beta$ and therefore it already represents the average total variance $\bar{V}_{\mathrm{post}}$ we are interested in determining.
    
    Moreover, it depends only on the variance $\sigma_0^{2}$ of the prior and the squeezing strength $r$ of the probe state. 
    For a fixed prior, the average posterior variance of both coordinates from Eq.~(\ref{var2_dis_het}) is minimized for $r=0$, that is, when there is no squeezing of the probe state. 
    We thus have 
    \begin{align}
        \bar{V}_{\mathrm{post}}(r) &\geq\,\bar{V}_{\mathrm{post}}(r=0)\,=\, \frac{2\nr\sigma_{0}^{2}}{1+2\nr\sigma_{0}^{2}}\,.
        \label{eq:heterodyne displacement estimation total average}
    \end{align}
    However, squeezing can help to reduce the variance in one coordinate, but this reduction comes at the cost of increasing the variance of the other coordinate with respect to the case where $r=0$. 
    Irrespective of the squeezing strength, we observe that the variances for both phase space coordinates decrease with respect to the prior, but only slightly. 
    When one is interested in reducing the variance in only one of the coordinates, say $\alpha_{\mathrm{R}}$, one may note that the variance decreases monotonically for increasing $r$. 
    Nevertheless, even as $r\rightarrow \infty$ the variance of the posterior is still bounded from below by $(\sigma_{0}^{-2}+4)^{-1}$. 
    This residual variance originates in the intrinsic uncertainty of the coherent-state basis associated with the POVM representing heterodyne detection. 
    That is, no matter which measurement outcome is obtained, the precision with which the parameter is identified is limited by the width of the variance of the coherent state corresponding to this outcome. 

    Although coherent states already minimize the product of uncertainties,
    one can overcome this limitation by considering measurement bases that consist of states with a lower variance in the desired parameter (e.g., in $\alpha_{\mathrm{R}}$) than that of a coherent state, at the expense of a larger variance in the respective other quadrature.  
    For instance, one may choose a basis of squeezed coherent states to reduce the uncertainty of the measurement basis in one coordinate. 
    In this regard, a homodyne measurement in the quadrature $\hat{q}$, which we will consider next, can be thought of as a limiting case of a measurement in a basis of infinitely squeezed coherent states.
%%%%%%%%%%%%%%%%%%%%%%%%%%%%%%%%%%%%%%%%%%%%%%%%%%%%%%%%%%%%%%

\subsection{Homodyne measurement}
    For homodyne detection with respect to the quadrature $\hat{q}$, the POVM is $\{\ket{q}\!\!\bra{q}\}_{q\in\mathbb{R}}$.
    As before, we begin by considering a squeezed vacuum state $\ket{\xi}$ as probe state to estimate the unknown displacement $\alpha$.  
    The prior distribution of $\alpha$ is again assumed to be Gaussian with mean $\alpha_{0}$ and variance $\sigma_0^2$.
    The probability to obtain outcome $q$ after a displacement $\alpha$ is given by
    \begin{align}
        p(q\nr|\alpha) 
        &=|\langle q|\hat{D}(\alpha)|\xi\rangle|^{2}
        =\frac{\exp\left[-\frac{2(\alpha_{\mathrm{R}}-\frac{q}{\sqrt{2}})^2}{\cosh{2 r}-\cos{\varphi}\, \sinh{2 r}}\right]}{
        \sqrt{\pi(\cosh{2 r}-\cos{\varphi}\, \sinh{2 r})}}.
    \end{align}
    Note that, here, the likelihood does not depend on the imaginary part $\alpha_{\mathrm{I}}$ of the displacement. 
    This is expected, since homodyne detection in one quadrature is completely `blind' to the orthogonal quadrature.  
    Therefore, the mean and variance for the imaginary part of the displacement parameter remain unchanged with respect to the prior, and we can focus entirely on the real part.
    
    \begin{figure*}[t]
        \centering
        \includegraphics[trim={0cm 0cm 0cm 0mm},clip, width=0.95\textwidth]{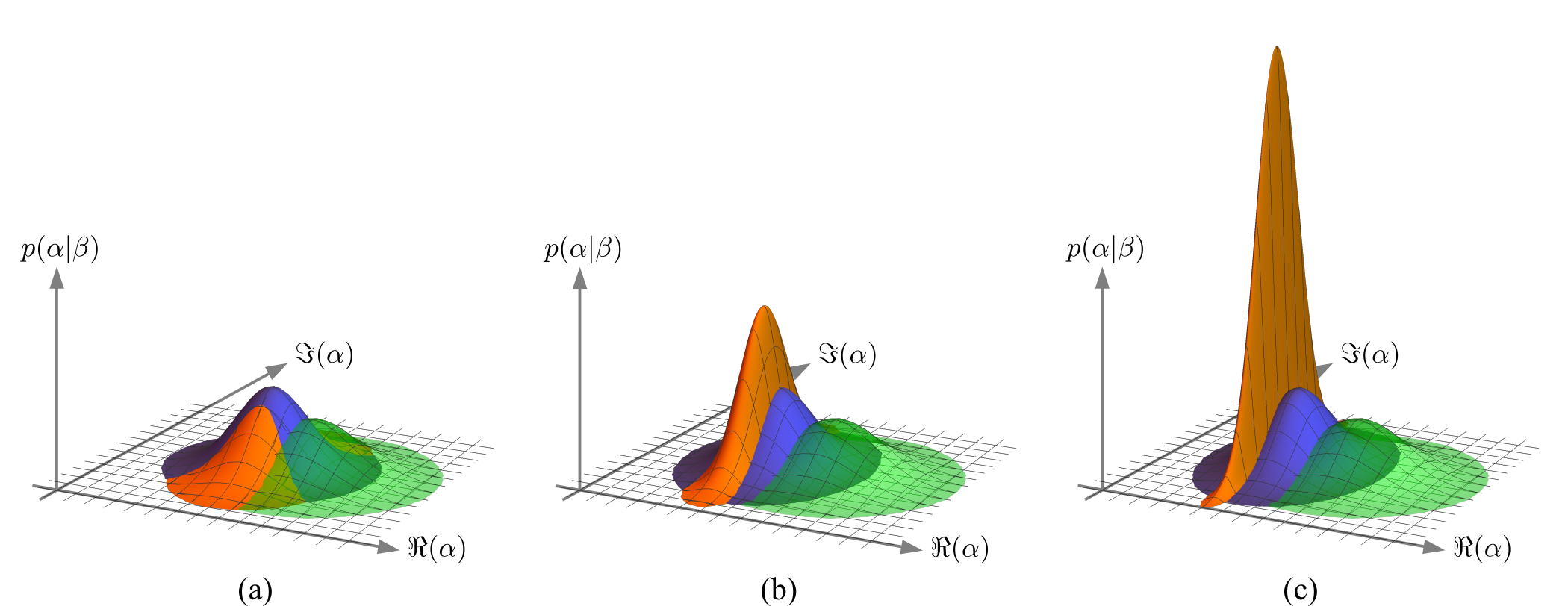}
        \caption{
        Displacement estimation using heterodyne and homodyne detection. The images show the same Gaussian prior (green) with initial standard deviation $\sigma_{0}=0.5$, and posterior distributions obtained for heterodyne (blue) and homodyne detection (orange) for different squeezing of the probe state, ranging from $r=0$ in (a), $r=1$ in (b), to $r=2$ in (c). The posterior distributions of the displacement parameter $\alpha$ given measurement outcome $q$ are Gaussian as well.
        } 
        \label{fig:homodyne_displ_est}
    \end{figure*}
    
    Since, once again the likelihood is a  Gaussian distribution in the measurement outcomes (here, in $q$), and thus proportional to a Gaussian distribution $\mathcal{N}_{\alpha_{\mathrm{R}}}(\langle\alpha_{\mathrm{R}}\rangle,\sigma^{2})$ in the estimated parameter with mean $\langle\alpha_{\mathrm{R}}\rangle=q/\sqrt{2}$ and variance $\sigma^2=(\cosh{2 r}-\cos{\varphi}\, \sinh{2 r})/4$, we can infer from Prop.~\ref{conjugatedprior} that the posterior is a Gaussian distribution with mean
    \begin{align}
        \hat{\alpha}_{\mathrm{R}} =\frac{2\sqrt{2}\,\sigma_0^2q+\alpha_{0,\mathrm{R}} (\cosh{2r}-\cos{\varphi}\,\sinh{2r})}{4\sigma_0^2+\cosh{2r}-\cos{\varphi}\, \sinh{2r}},
    \end{align}
    and variance
    \begin{align}
        \operatorname{Var}[p(\alpha_{\mathrm{R}}|q)]=\frac{\sigma_0^2(\cosh{2r}-\cos{\varphi}\,\sinh{2r})}{4\sigma_0^2+\cosh{2r}-\cos{\varphi}\ \sinh{2r}}.
    \end{align}

    The variance of the posterior distribution depends on the squeezing strength $r$ and the squeezing angle $\varphi$. 
    Both parameter hence provide room for optimization of the estimation procedure. 
    However, while increasing $r$ can be demanding experimentally and also comes at an increased energy cost for preparing the probe state, the relative angle $\varphi$ between the directions of measurement and squeezing can be varied freely without any particular practical or energetic restriction. 
    The variance is minimised for $\varphi=2\hspace*{0.5pt}n\pi$ and without loss of generality we choose $\varphi=0$. 
    For this choice, the average variance of the posterior for the chosen quadrature $\hat{q}$ is
    \begin{align}
        \bar{V}_{\mathrm{post}}^{\hat{q}} &=\,\operatorname{Var}[p(\alpha_{\mathrm{R}}|q)]\stackrel{\varphi=0}{=} \Bigl(\frac{1}{\sigma^2_0}+4\mathrm{e}^{2r}\Bigr)^{-1},
        \label{eq:minvar}
    \end{align}
    whereas the average total variance (again, for $\varphi=0$) is $\bar{V}_{\mathrm{post}}=\bar{V}_{\mathrm{post}}^{\hat{q}}+\sigma_{0}^{2}$. 
    Fig.~\ref{fig:homodyne_displ_est} shows a sample of different posterior distributions obtained by measurements with probe states with different squeezing. We observe that, whereas the marginal probability in $\hat{p}$ remains unchanged as the initial squeezing increases, the marginal probability in $\hat{q}$ becomes narrower. 
    We further note that for $r=0$ we recover the results obtained by Personick~\cite{Personick1971}.
%%%%%%%%%%%%%%%%%%%%%%%%%%%%%%%%%%%%%%%%%%%%%%%%%%%%%

\subsection{Comparison of measurement strategies}
    Let us now interpret and compare the results for Gaussian displacement estimation with heterodyne and homodyne measurements. 
    For homodyne detection, squeezing in the probe state results in an average posterior variance in $\hat{q}$, given by Eq.~(\ref{eq:minvar}), that rapidly decreases to $0$ as the squeezing strength $r$ increases. 
    While the posterior variance in $\hat{q}$ can thus be arbitrarily close to zero in the homodyne detection scenario, this comes at the cost of not reducing the variance in $\hat{p}$ at all. 
    We thus have $\lim_{r\rightarrow\infty}\bar{V}\suptiny{0}{0}{\mathrm{homodyne}}_{\mathrm{post}}=\sigma_{0}^{2}$. Comparing this with the result for heterodyne detection in Eq.~(\ref{eq:heterodyne displacement estimation total average}), we see that $\bar{V}\suptiny{0}{0}{\mathrm{homodyne}}_{\mathrm{post}}\geq \bar{V}\suptiny{0}{0}{\mathrm{heterodyne}}_{\mathrm{post}}(r=0)$ for priors with variance $\sigma_{0}^{2}\geq 1/2$, independently of the squeezing strength used with the homodyne detection. 
    However, for more narrow priors, homodyne detection supplemented by squeezed probe states can outperform heterodyne detection in terms of the total variance only if the squeezing is strong enough, i.e., when $r>-\tfrac{1}{2}\ln(1-2\sigma_{0}^{2})$. 
    
    However, when we focus on the estimation of only one of the quadratures, here quadrature $\hat{q}$, then homodyne detection outperforms heterodyne detection for all prior widths and for all squeezing strengths, even if different squeezing strengths are compared for the two detection methods. That is, the limit of $r\rightarrow\infty$ for heterodyne detection in Eq.~(\ref{eq:var_dis_het}) coincides with the homodyne detection case where $r=0$ in Eq.~(\ref{eq:minvar}), and we thus find
    \begin{align}
        \bar{V}\suptiny{0}{0}{\hat{q},\mathrm{homodyne}}_{\mathrm{post}}
        \,\leq\,
        \frac{\sigma_{0}^{2}}{1+4\sigma_{0}^{2}}
        \,\leq\,
        \bar{V}\suptiny{0}{0}{\hat{q},\mathrm{heterodyne}}_{\mathrm{post}}.
    \end{align}
    
    We can also compare these results to more general measurement strategies. 
    For a Gaussian prior (in a single parameter), the Fisher information of the prior (see Sec.~\ref{sec:quantum estimatio in Bayesian setting}) evaluates to $I[p(\alpha_{\mathrm{R}})]=1/\sigma_0^2$. 
    At the same time, the QFI for a single-mode Gaussian state is bounded by
    $\mathcal{I}(\rho)\le 4\mathrm{e}^{2r}$ (cf. Eq.~(15) and subsequent text in Ref.~\cite{PinelJianTrepsFabreBraun2013}). 
    With this, the Van Trees inequality in the form of Eq.~(\ref{eq:QBCR_bound}) reads
    \begin{align}
        \bar{V}\suptiny{0}{0}{\hat{q}}_{\mathrm{post}}
        \ge\left(\frac{1}{\sigma^2_0}+4\mathrm{e}^{2r}\right)^{-1}.
    \end{align}
    This shows that the combination of single-mode squeezing and homodyne detection is the optimal strategy for Bayesian estimation of one coordinate of displacement (or displacement radius with known phase) with a single-mode Gaussian probe state.

    Finally, let us consider repeated measurements, which can easily be accommodated within the framework of conjugate priors. 
    In particular, we know that the posterior is of the same form as the prior, i.e., both are normal distributions. Since the posterior distribution is used as the prior for the next measurement round, we obtain a recursive formula for the average variance, given by
    \begin{align}
        \sigma_{m+1}^2=\frac{\sigma_m^2 \operatorname{Var}[p(q|\alpha)]}{\sigma_m^2 + \operatorname{Var}[p(q|\alpha)]},
    \end{align}
    where $\sigma_m$ is the variance of round $m$. 
    Since $\operatorname{Var}[p(q|\alpha)]=\mathrm{e}^{-2r}/4$ depends only on the squeezing of the probe state, this term is constant for the same probe state. 
    Solving the recursive equation gives
    \begin{align}
        \sigma_{m}^2= \left(\frac{1}{\sigma_0^2}+4\nr m\, \mathrm{e}^{2r}\right)^{-1}.
    \end{align}
    Moreover, we note that repeated measurements include the possibility of a sequential measurement strategy that provides information about both components of the displacement. 
    For instance, the squeezing in the probe states and the direction of the homodyne measurement can be tailored towards estimating the real part in one half of the estimation rounds, while the remaining rounds are used to estimate the imaginary part. 
    We conclude this section by noting that already a quite simple setup, consisting of (limited) squeezing in the probe states combined with homodyne detection, can provide accurate information for Bayesian estimation of displacements. 

%%%%%%%%%%%%%%%%%%%%%%%%%%%%%%%%%%%%%%%%%%%%%%%%%%%%%%%%%%%%%%%%%%%%%%%%%%%%%%%%%%%%%%%

\section{Phase estimation} 
\label{sec:phase}
    We now come to the paradigmatic case of phase estimation, which we want to examine within the framework of Bayesian estimation using Gaussian states and measurements. 
    Historically, phase estimation has been closely associated with interferometry~\cite{Michelson1887}, but nowadays, phase estimation is usually considered in a broader context.
    In particular, Bayesian phase estimation has been studied for a variety of applications, see, e.g.,~\cite{Wiebe2016, Paesani2017, Martinez2019}.
    While there are some studies identifying optimal estimation strategies using Gaussian states and measurements~\cite{Oh2019a,Oh2019b,Ataman2019}, these operate within the local estimation paradigm and hence fall outside of the Bayesian phase estimation framework we consider here. 
    We therefore focus on a special case of Bayesian phase estimation, where there is no prior information on the phase and local estimation hence cannot be employed in a meaningful way.
    For such cases, we wish to identify simple strategies based on Gaussian states and measurements that can efficiently narrow the prior down to the point where local estimation can take over. 

    Specifically, we consider a phase estimation scenario where a phase rotation operator as in Eq.~\eqref{eq:phase rotation} is applied to a single-mode Gaussian probe state.  
    We consider the phase $\theta\in[-\pi,\pi)$ to be entirely unknown initially, such that the prior is a uniform distribution on the chosen interval, i.e., $p(\theta)=1/2\pi$. 

    In the following sections, we then study the performance of heterodyne and homodyne detection in this estimation scenario, and we adapt the specific probe states to the respective measurements. 
    In particular, we note that, although the optimal probe state (at fixed average energy) for local phase estimation is a single-mode squeezed state, this is not necessarily the case for Bayesian estimation.

\subsection{Heterodyne measurement}
\label{sec:phase_hetero}
    For Gaussian phase estimation with heterodyne measurements, we consider probe states that are squeezed with strength $r=|\xi|$ before being displaced, i.e., probe states of the form $\hat{D}(\alpha)\hat{S}(r \mathrm{e}^{i\varphi})\ket{0}$, where $r \ge 0$ and $\varphi \in [0, 2\pi)$.
    Whereas the most general Gaussian single-mode probe states are determined by arbitrary complex values $\alpha$ and $\xi$, i.e., displacement and squeezing with arbitrary strength along arbitrary directions, the rotational symmetry of the phase estimation problem with heterodyne measurements allows one to fix one of these directions.
    Without loss of generality, we therefore choose $\alpha=|\alpha|$ to be real and positive.
    More specifically, we assume that the displacement is strictly non-zero, $\alpha>0$, since the vacuum state is rotationally invariant, and not even a squeezed vacuum state can be used to distinguish between rotations around $\theta$ and $\theta+\pi$. 

    For the squeezing direction, it is then quite intuitive to see that squeezing along the quadrature $\hat{p}$ ($\varphi=\pi$, $\xi=-r<0$) is optimal for single-mode phase estimation when $\alpha>0$ and when heterodyne measurements are used.
    That is, when the variance of the Gaussian state is initially reduced along the quadrature $\hat{p}$, the Wigner function becomes concentrated along the $\hat{q}$-quadrature, decreasing the variance in the phase of the initial state, and hence also decreasing the variance in the phase of the encoded state $\rho(\theta)$.
    When applying the heterodyne measurement, the probability for obtaining an outcome $\beta$ whose phase matches the unknown phase $\theta$ is thus increased.
    Conversely, probe states that are squeezed along the same direction as the initial displacement have an increased phase variance and are therefore less useful for phase estimation.
    In the remainder of this section, we therefore focus on probe states of the form $\hat{D}(\alpha)\hat{S}(-r)\ket{0}$. 

    However, since the calculations and results for arbitrary values of $r$ are still quite unwieldy, we first consider the simple case where the probe state is not squeezed at all but just a coherent state $\ket{\alpha}$ (Sec.~\ref{sec:coherent probe and heterodyne}).
    Then we present the results for squeezing along the optimal direction, $\xi=-r<0$, with respect to the displacement $\alpha>0$ (Sec.~\ref{sec:displaced squeezed probe and heterodyne}).
%%%%%%%%%%%%%%%%%%%%%%%%%%%%%%%%%%%%%%%%%%%%%%%%%%%%%%%%

\subsubsection{Coherent states \& heterodyne detection}
\label{sec:coherent probe and heterodyne}
    Here, the probe state is $\ket{\alpha}$ with $\alpha>0$. 
    The action of the phase rotation operator $\hat{R}(\theta)$ [Eq.~\eqref{eq:phase rotation}] results in the encoded state $\hat{R}(\theta)\ket{\alpha}=\ket{\mathrm{e}^{-i\theta}\alpha}$. The likelihood to obtain outcome $\beta\in\mathbb{C}$, given that the phase has the value $\theta$, is given by
    \begin{align}
        p(\beta\nr|\nr\theta) &=\,
        \tfrac{1}{\pi}|\scpr{\beta}{\mathrm{e}^{-i\theta}\alpha}|^{2}
        \,=\,
        \tfrac{1}{\pi} \mathrm{e}^{-|\mathrm{e}^{i\theta}\beta-\alpha|^{2}}.
    \end{align}
    Writing $\beta=|\beta|\mathrm{e}^{-i\phi_{\beta}}$ and  $|\mathrm{e}^{i\theta}\beta-\alpha|^{2}=\alpha^{2}+|\beta|^{2}-2\alpha|\beta|\cos(\theta-\phi_{\beta})$, we can express the (unconditional) probability to obtain outcome $\beta$ as
    \begin{align}
        p(\beta)    &=\,
        \int\limits_{-\pi}^{\pi}\!\!\!d\theta\,p(\theta)\,
        p(\beta\nr|\nr\theta)\,=\,
        \frac{\mathrm{e}^{-(\alpha^{2}+|\beta|^{2})}}{\pi}\,I_{0}(2\alpha|\beta|),  
        \label{eq:p of beta}
    \end{align}
    where $I_{0}(x)$ is the modified Bessel function of the first kind. Using Bayes' law, the posterior is given by
    \begin{align}
        p(\theta\nr|\nr\beta)   &=\,
        \frac{p(\theta)\,p(\beta\nr|\nr\theta)}{p(\beta)}
        \,=\,
        \frac{\mathrm{e}^{2\alpha|\beta|\cos(\theta-\phi_{\beta})}}{2\pi\,I_{0}(2\alpha|\beta|)}.    
    \end{align}
    Since we are considering a parameter with a range whose endpoints $\pm\pi$ are identified, it is useful to consider estimators and variances that are invariant under shifts by $2\pi$. For the estimator we therefore choose $\hat{\theta}(\beta)=\arg\langle \mathrm{e}^{i\theta}\rangle_{p(\theta\nr|\nr\beta)}$. 
    As we discuss in more detail in Appendix~\ref{sec:app_phase_estimation_heterodyne}, the estimator evaluates to
    \begin{align} \label{eq:estimator_coherent_hetero_phase}
        \hat{\theta}(\beta)
        &=\,
        \arg\Bigl[
        \int\limits_{-\pi}^{\pi}\!\!\! d\theta\,
        p(\theta\nr|\nr\beta) \mathrm{e}^{i\theta}
        \Bigr]
        \,=\,
        \phi_{\beta},
    \end{align}
    and hence corresponds to the phase $\phi_{\beta}$ of the measurement outcome $\beta$. 

    To evaluate the performance of this estimation strategy, we calculate the average variance of the posterior as done in the above sections.
    However, instead of an expression such as in Eq.~(\ref{eq:posterior variance MSE}), we now use a covariant variance that is invariant under shifts by $2\pi$, by taking the average of $\sin^2\bigl[\theta - \hat{\theta}(\beta)\bigr]$ rather than of $\left(\theta - \hat{\theta}(\beta)\right)^2$.\footnote{We note here that the chosen variance is invariant also under shift of the estimator by integer multiples of $\pi$, not just shift by even multiples of $\pi$. In principle, one could also use quantifiers for the width of the distribution that depend only on $|\langle \mathrm{e}^{i\theta}\rangle_{p(\theta\nr|\nr\beta)}|$, such as the Holevo phase variance~\protect\cite{Holevo1984}, which are completely independent of the value of the estimator.
    The choice we make here is motivated by the better comparison with the homodyne detection scenario in Sec.~\protect\ref{sec:phase estimation homodyne}, where the phase can only be resolved within an interval of length $\pi$.
    }
    Specifically, we calculate
    \begin{align}
        V_{\mathrm{post}}(\beta) &=\!
        \int\limits_{-\pi}^{\pi}\!\!\! d\theta\,
        p(\theta\nr|\nr\beta)\,
        \sin^2\bigl[\theta - \hat{\theta}(\beta)\bigr]=
        \tfrac{\,_{0}F_{1}(2;\alpha^{2}|\beta|^{2})}{2\,I_{0}(2\alpha|\beta|)\,\Gamma(2)},
        \label{eq:posterior variance phase est heterodyne}
    \end{align}
    where $\,_{0}F_{1}(a;z)$ is the confluent hypergeometric function and $\Gamma(z)$ is the Euler gamma function.
    Despite the complicated form of the posterior and the variance, the average variance then simply becomes
    \begin{align} 
        \bar{V}_{\mathrm{post}}
        &=\int\!\!\! d^{2}\nl\beta\ p(\beta)\,V_{\mathrm{post}}(\beta)\,=\,
        \frac{1-\mathrm{e}^{-|\alpha|^2}}{2\,|\alpha|^2},
        \label{eq:ageV_phase_displ}
    \end{align}
    as we discuss in more detail in Appendix~\ref{sec:app_phase_estimation_heterodyne}. 
    In terms of the average photon number $n=\left|\alpha\right|^2$, which is proportional to the average energy of the probe state, the average variance of the posterior hence scales as $1/{n}$ as $n\to\infty$, as can be expected for `classical' probe states such as the coherent states considered here. 
%%%%%%%%%%%%%%%%%%%%%%%%%%%%%%%%%%%%%%%%%%%%%%%%%%%%%

\subsubsection{Displaced squeezed states \& heterodyne detection}
\label{sec:displaced squeezed probe and heterodyne}
    Let us now consider probe states that are squeezed with strength $r$ before being displaced, i.e., probe states of the form $\hat{D}(\alpha)\hat{S}(-r)\ket{0}$, where we assume $\alpha,r\in\mathbb{R}$ with $\alpha>0$ and $r>0$ as mentioned. 
    For the heterodyne measurement, the likelihood to obtain outcome $\beta$ given the phase $\theta$ is given by
    \begin{align} \label{eq:likelihood_phase_squeezed}
        p(\beta\nr|\nr\theta)
        &=\tfrac{1}{\pi}|\bra{\beta}\hat{R}(\theta)\hat{D}(\alpha)\hat{S}(-r)\ket{0}|^{2}
        \nonumber\\
        &=\tfrac{1}{\pi}\mathcal{F}\bigl(\ket{\mathrm{e}^{i\theta}\beta},\ket{\alpha,-r}\bigr).
    \end{align}
    For the fidelity of the two Gaussian states, we can again refer to Eq.~(\ref{eqfidelity}), where $\rho_{1}=\ket{\mathrm{e}^{i\theta}\beta}\!\!\bra{\mathrm{e}^{i\theta}\beta}$ and $\rho_{2}=\ket{\alpha,-r}\!\!\bra{\alpha,-r}$, for which the first moments are
    \begin{equation*}
        \mathbf{\bar{x}}_{1}=\mathbf{\bar{x}}_{\mathrm{e}^{i\theta}\beta}=\sqrt{2}\begin{pmatrix} \Re(\mathrm{e}^{i\theta}\beta)\\ \Im(\mathrm{e}^{i\theta}\beta)\end{pmatrix}
        \ \  \text{and}\ \ 
        \mathbf{\bar{x}}_{2}=\mathbf{\bar{x}}_{\alpha,-r}=\sqrt{2}\begin{pmatrix} \Re(\alpha)\\ \Im(\alpha)\end{pmatrix}.
    \end{equation*}
    The second moments of these states are represented by
    \begin{equation*}
        \mathbf{\Gamma}_{1}=\mathbf{\Gamma}_{\mathrm{e}^{i\theta}\beta}=\mathds{1}_{2} 
        \ \ \ \  \text{and}\ \ \ \ 
        \mathbf{\Gamma}_{2}=\mathbf{\Gamma}_{\alpha,-r}=\begin{pmatrix}\mathrm{e}^{2r} & 0\\ 
            0 & \mathrm{e}^{-2r}\end{pmatrix},
    \end{equation*}
    respectively.
    Since $\det\Gamma_{1}=\det\Gamma_{2}=1$ and $\det(\Gamma_{1}+\Gamma_{2})=4\cosh^{2}(r)$, we then have
    \begin{align} \label{eq:likelihood_phase_squeezed_result}
        p(\beta\nr|\nr\theta)
        &=\,
        \frac{\exp\Bigl[-\frac{\mathrm{e}^{-r}\Re^{2}(\mathrm{e}^{i\theta}\beta-\alpha)+
        \mathrm{e}^{r}\Im^{2}(\mathrm{e}^{i\theta}\beta-\alpha)}{\cosh r}\Bigr]}{\pi\cosh r}.
    \end{align}
    As we explain in more detail in Appendix~\ref{sec:app_phase_estimation_heterodyne_coherent_squeezed_probe}, the (unconditional) probability to obtain outcome $\beta$ can then be written as an infinite sum of Bessel functions of the first kind by using the Jacobi-Anger expansion, which results in 
    \begin{align}
        p(\beta)    &=\,\frac{\mathrm{e}^{-\alpha^2(1-\tanh r)-|\beta|^2}}{\pi\nr\cosh r}
        \sum\limits_{\substack{m_{1},m_{2}\\ =-\infty}}^{\infty}
        \mathrm{e}^{-i\nr m_{1}\nr\pi}\,I_{-2m_{1}-m_{2}}(-2\alpha|\beta|)\nonumber\\[1mm]
        &\ \ \ \times\,I_{m_{1}}(-|\beta|^{2}\tanh r)\,I_{m_{2}}(2\alpha|\beta|\tanh r).
        \label{eq:pbeta_phase_squdis}
    \end{align}
    Using Bayes' law, the posterior can then be obtained directly as $p(\theta\nr|\nr\beta)=p(\beta\nr|\nr\theta)/\bigl[2\pi\,p(\beta)\bigr]$ with the likelihood from Eq.~(\ref{eq:likelihood_phase_squeezed_result}) and $p(\beta)$ as in Eq.~(\ref{eq:pbeta_phase_squdis}). 
    Similarly, we can use the Jacobi-Anger expansion to evaluate $\left\langle \mathrm{e}^{i\theta}\right\rangle=\int_{-\pi}^{\pi}\! d\theta\, p(\theta\nr|\nr\beta) \mathrm{e}^{i\theta}$.
    As shown explicitly in Appendix~\ref{sec:app_phase_estimation_heterodyne_coherent_squeezed_probe}, one finds $\Im\bigl(\left\langle \mathrm{e}^{i(\theta-\phi_{\beta})}\right\rangle\bigr)=0$, and the estimator is hence given by
    \begin{align}
        \hat{\theta}(\beta) &=
        \phi_{\beta}\ \text{or}\ \phi_{\beta}+\pi, 
    \end{align}
    i.e., the estimate either corresponds to the phase $\phi_{\beta}$ of the measurement outcome $\beta$, or is shifted by $\pi$. 
    
\begin{figure}[ht!]
       \centering
       \includegraphics[width=0.95\columnwidth]{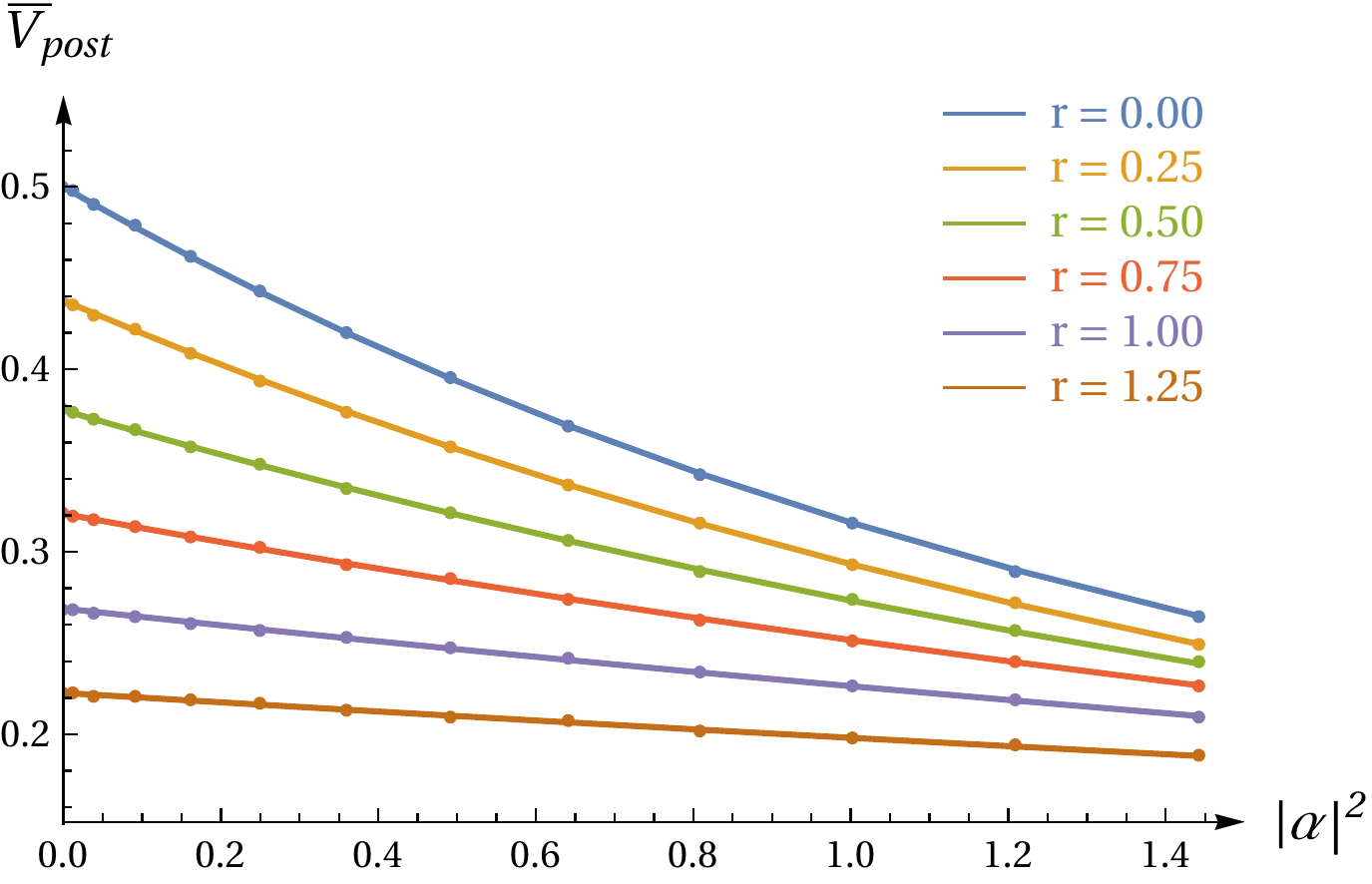}\\
       (a)\\
       \includegraphics[width=0.95\columnwidth]{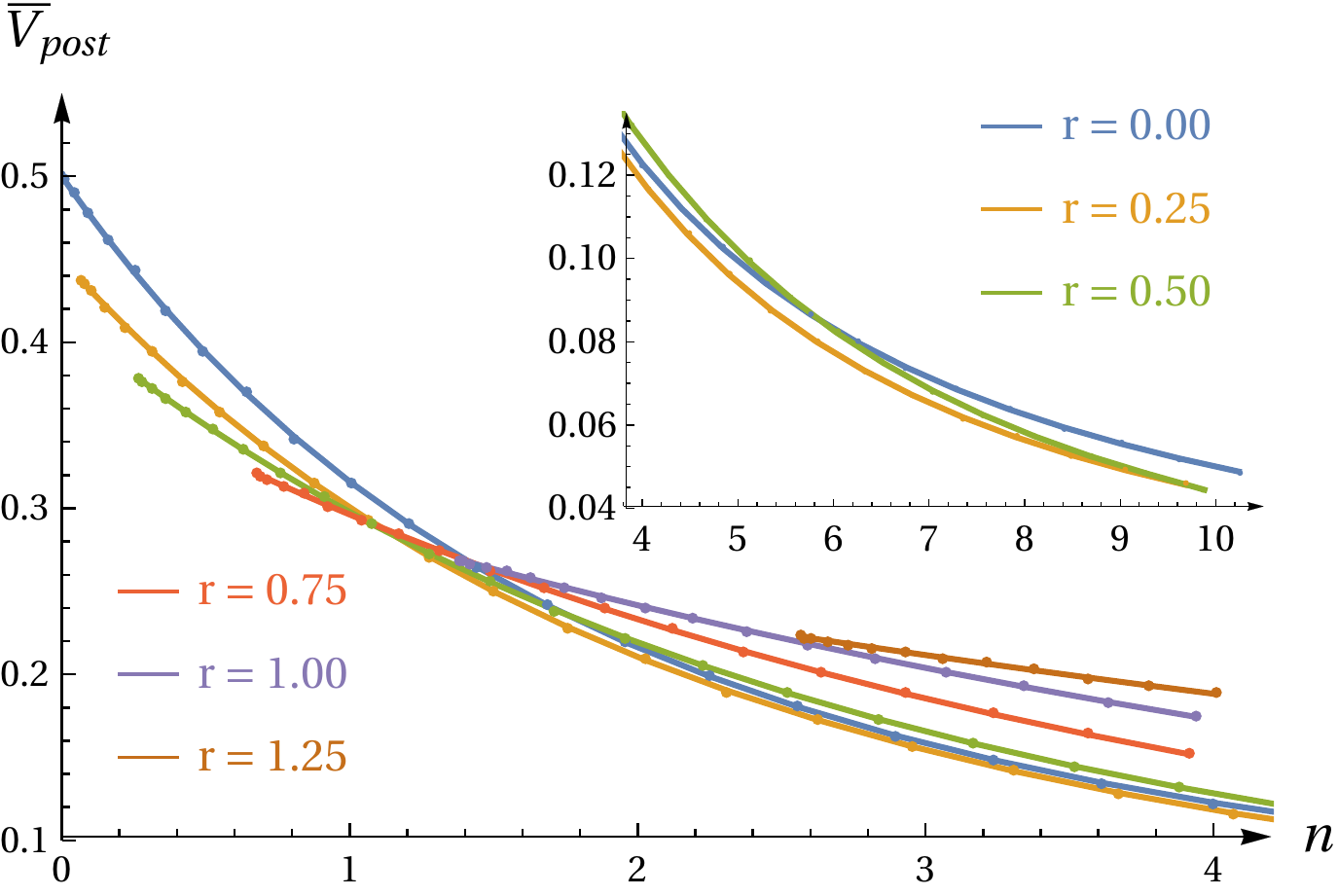}\\
       (b)
       \caption{
       Bayesian phase estimation with single-mode Gaussian probes and heterodyne measurements. 
       (a) The average variance $\bar{V}_{\mathrm{post}}$ from Eq.~(\protect\ref{eq:aveV_phase_squedisp}) is shown for different values of $\alpha\geq0$ and $r\geq0$ as a function of $\alpha^{2}$. The line on the top represents the average variance for purely displaced probe states ($r=0$) from Eq.~\protect\eqref{eq:ageV_phase_displ}.
       The lines below indicate results of numerically evaluating Eq.~(\protect\ref{eq:aveV_phase_squedisp}) for different values of $\alpha$ for fixed values of $r$ from $r=0.25$ to $r=1.25$ (top to bottom, starting at the second line from the top).
       (b) The average variance $\bar{V}_{\mathrm{post}}$ is shown as a function of the average photon number $n=|\alpha|^2+\sinh^2{r}$. 
       The lines do not start at $n=0$ because the nonzero values of $r$ give rise to non-zero average energies even for $\alpha=0$. The inset shows how the lines for $r=0$, $r=0.25$, and $r=0.5$ continue as $n$ increases. 
       }
       \label{fig:phase_hetero}
    \end{figure}

    To see if squeezing improves the estimation, we calculate the variance of the posterior, $V_{\mathrm{post}}(\beta) =\int_{-\pi}^{\pi}\! d\theta\, p(\theta\nr|\nr\beta)\, \sin^2[\theta - \hat{\theta}(\beta)]$, and its average, and compare the latter with the corresponding value obtained for coherent probe states.
    Specifically, we obtain the expression (see Appendix~\ref{sec:app_phase_estimation_heterodyne_coherent_squeezed_probe} for more details)
    \begin{align} 
        \bar{V}_{\mathrm{post}}
        &
    =
        \tfrac{\mathrm{e}^{-\alpha^2(1-\tanh{r})}}{\cosh r}
        \sum_{\substack{n_2,n_3\\ =-\infty}}^{\infty}
        \int\limits_{0}^{\infty}\!\!\! d|\beta|\,
        |\beta|
        \, \mathrm{e}^{-|\beta|^2}
        \,I_{n_2}(-|\beta|^2\tanh{r})
        \nonumber\\
        &
        \ 
        \times 
        I_{n_3}(2\alpha|\beta|\tanh{r})
        \tfrac{1}{2}(-1)^{n_2}\Bigl[
        2I_{-2n_2-n_3}(-2\alpha\left|\beta\right|)
        \nonumber\\[1mm]
        &
        \ 
        -I_{2-2n_2-n_3}(-2\alpha\left|\beta\right|)
        -I_{-2-2n_2-n_3}(-2\alpha\left|\beta\right|)
        \Bigr].
        \label{eq:aveV_phase_squedisp}
    \end{align}
    Unfortunately, the analytical solution of the integral and double-sum in Eq.~(\ref{eq:aveV_phase_squedisp}) is unknown.
    We have therefore numerically evaluated the average variance $\bar{V}_{\mathrm{post}}$ for different values of $\alpha$ and $r$.
    As illustrated by the sample plots in Fig.~\ref{fig:phase_hetero}~(a), for any fixed displacement, squeezing improves the estimation precision as measured by the average variance beyond the value achievable by displacements alone, where the latter is represented by Eq.~(\ref{eq:ageV_phase_displ}).
    This is in agreement with the intuition provided by the Wigner function of the probe states: Squeezing along the $\hat{p}$-quadrature ($\xi=-r<0$) of a coherent state displaced along the $\hat{q}$ axis ($\alpha>0$) leads to a concentration of the Wigner function around the $\hat{q}$-axis, that is rotated around the origin by the phase rotation, visually resembling a clock dial.
    Increased squeezing narrows the width of this `dial', making it more likely to obtain measurement outcomes $\beta$ whose phase matches the phase to be estimated.
    
    However, when considering constraints on the average energy of the probe state, here represented by the average photon number $n=|\alpha|^2+\sinh^2{r}$, squeezing is only beneficial in certain regimes.
    For relatively strong squeezing such as $r=1$ or $r=1.25$, the average variance is larger for squeezed-displaced states than for purely displaced states with the same average photon number, as illustrated in Fig.~\ref{fig:phase_hetero}~(b).
    This can be understood from the fact that the average photon number required for a squeezing of $r=1.25$ is sufficient for a coherent state that is displaced more than 2 standard deviations from the origin and hence already provides a clear phase reference. 
    For smaller squeezing, such as for $r=0.75$, there is a regime of small photon numbers where the combination of squeezing and displacement can outperform pure displacement.
    This can also be readily understood, while such a squeezed vacuum state already has a standard deviation $\Delta\hat{p}\nr(|\xi\rangle)=\mathrm{e}^{-r}/\sqrt{2}$
    less than half of that of a coherent state, a coherent state with the same average energy is displaced by only $\sqrt{2}\alpha=\sqrt{2\sinh^2{r}}\approx1.64\Delta\hat{q}(|\alpha\rangle)$.
    However, for larger $n$ (already around $n\approx1.41$) pure displacements become better, see Fig.~\ref{fig:phase_hetero}~(b). 
    Finally, we see that for even smaller values of $r$, such as for $r=0.5$, there is only a specific range of values for $n$ where purely coherent probes are more efficient, while low squeezing ($r=0.25$) added to the displacement outperforms pure displacement for the entire range of $n$ that we have explored numerically.
    At the same time, in terms of the difference between the average variances achieved, e.g., for $r=0$ and $r=0.25$, the advantage obtained from using a slightly squeezed state seems to be at least an order of magnitude smaller than the average variances achieved (in the explored parameter range). 
%%%%%%%%%%%%%%%%%%%%%%%%%%%%%%%%%%%%%%%%%%%%%%%%%%%%%%%%%%%%%%%%%%%%%%%%%%

\subsection{Homodyne measurement}\label{sec:phase estimation homodyne}
    Here, we consider Bayesian phase estimation with single-mode Gaussian probe states combined with homodyne measurements in the quadrature $\hat{q}$.
    Since this kind of measurement provides no information on the complementary quadrature $\hat{p}$, it cannot distinguish between phases phases of $\theta$ and $-\theta$. 
    Thus, we restrict the range of $\theta$ to $[0,\pi]$, and the prior distribution is given by $p(\theta) = 1/\pi$.
    
    \subsubsection{Coherent states \& homodyne detection}\label{sec:coherent probe and homodyne}
    
    As before in Sec.~\ref{sec:phase_hetero}, we start with the case where the probe state is a coherent state, $\hat{D}(\alpha)\ket{0}=\ket{\alpha}$ for $\alpha>0$. 
    The likelihood to obtain outcome $q\in\mathbb{R}$ can be written as
    \begin{align} \label{eq:likelihood_coheret_homo_phase 1}
        p(q|\theta)
        &=\,
        |\langle q| \mathrm{e}^{-i\theta} \alpha \rangle|^{2}
        \,=\,\int\limits_{-\infty}^{\infty}\!\!\!dp\,W(q,p)\,, 
    \end{align}
    where $W(q,p)$ is the Wigner function of the rotated coherent state $\ket{\mathrm{e}^{-i\theta}\alpha}$.
    The latter can be obtained from Eq.~(\ref{eq:wignerfcn}) by noting that $\mathbf{\Gamma}_{\mathrm{e}^{-i\theta}\alpha}=\mathds{1}_{2}$ and $\mathbf{\bar{x}}_{\mathrm{e}^{-i\theta}\alpha}=\sqrt{2}\alpha(\cos\theta,-\sin\theta)^{T}$.
    With this, one finds that 
    \begin{align} \label{eq:likelihood_coheret_homo_phase}
        p(q|\theta)
        &=\,
        \tfrac{1}{\sqrt{\pi}}
        \mathrm{e}^{-
        \left(q -\sqrt{2}\alpha\cos{\theta}\right)^2}.
    \end{align}
    Further noting that the range of $\theta$ is $[0,\pi]$, the (unconditional) probability to obtain $q$ can be expressed as 
    \begin{align}
        p(q)
        &=
        \int_{0}^{\pi}d\theta \, p(\theta) \, p(q|\theta)
        \nonumber\\
        &=
        \tfrac{1}{\sqrt{\pi}} \mathrm{e}^{-q^2-\alpha^2}
        \sum_{m=-\infty}^{\infty} I_{2m}(2\sqrt{2}q\alpha) I_m(-\alpha^2),
        \label{eq:phase homodyne unconditonal}
    \end{align}
    as we show in detail in Appendix \ref{sec:app_phase_estimation_homodyne_coherent_probe}. 
    Using Bayes' law, the posterior $p(\theta\nr|\nr q)=p(q\nr|\nr\theta)/\bigl[\pi\,p(q)\bigr]$ is then just obtained by inserting $p(q|\theta)$ and $p(q)$ from Eqs.~(\ref{eq:likelihood_coheret_homo_phase 1}) and~(\ref{eq:phase homodyne unconditonal}), respectively.  
    In Appendix \ref{sec:app_phase_estimation_homodyne_coherent_probe} we also explicitly calculate the circular moment, which we find to be given by
    \begin{align}
        \langle \mathrm{e}^{i\theta} \rangle
        &=
        \tfrac{1}{M}\,
        \sum\limits_{n=-\infty}^{\infty}
        I_{2n+1}(2\sqrt{2}q\alpha)
        I_{n}(-\alpha^2)
        \\
        &\ 
        +
        i
        \tfrac{2}{M\pi}\!\!
        \sum\limits_{m,n=-\infty}^{\infty}\!\!
        \tfrac{
        I_m(-\alpha^2)
        I_{2n}(2\sqrt{2}q\alpha)
        (1-4m^2-4n^2)
        }{
        (2n-2m-1)(2n-2m+1)(2n+2m+1)(2n+2m-1)
        }
        ,\nonumber
    \end{align}
    where 
    \begin{align}
        M
        &:=
        \sum_{m=-\infty}^{\infty}
        I_{2m}(2\sqrt{2}q\alpha)
        I_m(-\alpha^2)\,.
    \end{align}
    As we see, already the expression for the estimator $\hat{\theta}(q)=\arctan\bigl[\Im(\langle \mathrm{e}^{i\theta} \rangle)/\Re(\langle \mathrm{e}^{i\theta} \rangle)\bigr]$ for a coherent probe state is sufficiently more complicated than its counterpart in the case of heterodyne measurements [cf.~Eq.~(\ref{eq:estimator_coherent_hetero_phase})].
    We therefore resort to a numerical evaluation of the variance $V_{\mathrm{post}}(q)= \int_{0}^{\pi}\! d\theta\, p(\theta\nr|\nr q)\, \sin^2\bigl[\theta - \hat{\theta}(q)\bigr]$ and the average variance $\bar{V}_{\mathrm{post}} =\int_{-\infty}^{\infty}\! dq\, p(q)\,V_{\mathrm{post}}(q)$ already for the case of coherent probe states.
    The results for $\bar{V}_{\mathrm{post}}$ as a function of $n=\alpha^{2}$ are shown in Fig.~\ref{fig:phase_homo}, together with the corresponding average variance for squeezed probe states, which we will briefly discuss next. 
%%%%%%%%%%%%%%%%%%%%%%%%%%%%%%%%%%%%%%%%%%%%%%%%%%%%%%%%%%%%%%%%%%%%%%%%%

\subsubsection{Displaced squeezed states \& homodyne detection}
    In the present section, we consider a squeezed and displaced probe state, $\hat{D}(\alpha)\hat{S}(r \mathrm{e}^{i\varphi})\ket{0}$ for $\alpha\geq0$ and $\varphi\in [0,2\pi)$. 
    While the optimal squeezing angle for heterodyne measurements is $\varphi=\pi$, the optimal $\varphi$ for homodyne measurements depends on the phase $\theta$. 
    
    Since the homodyne measurement informs us of the value of the quadrature $\hat{q}$, the squeezing direction of the probe state is optimal, when the rotated probe state $\hat{R}(\theta)\hat{D}(\alpha)\hat{S}(r \mathrm{e}^{i\varphi})\ket{0}=\hat{R}(\theta)\hat{D}(\alpha)\hat{R}(\varphi/2)\hat{S}(r)\ket{0}$ is squeezed along the $\hat{q}$-quadrature such that its Wigner function is elongated along the $\hat{p}$-quadrature.
    Thus, for any fixed $\theta$, the optimal squeezing angle satisfies $\theta+\tfrac{\varphi}{2}=m\pi$ for $m\in\mathbb{Z}$, i.e. $\varphi=2(m\pi-\theta)$.
    However, since we consider a flat prior and there is hence no initial information on $\theta$ available, we leave the squeezing angle as a variable for the following calculations.
    
\begin{figure}[t!]
    \centering
        \includegraphics[width=0.97\columnwidth]{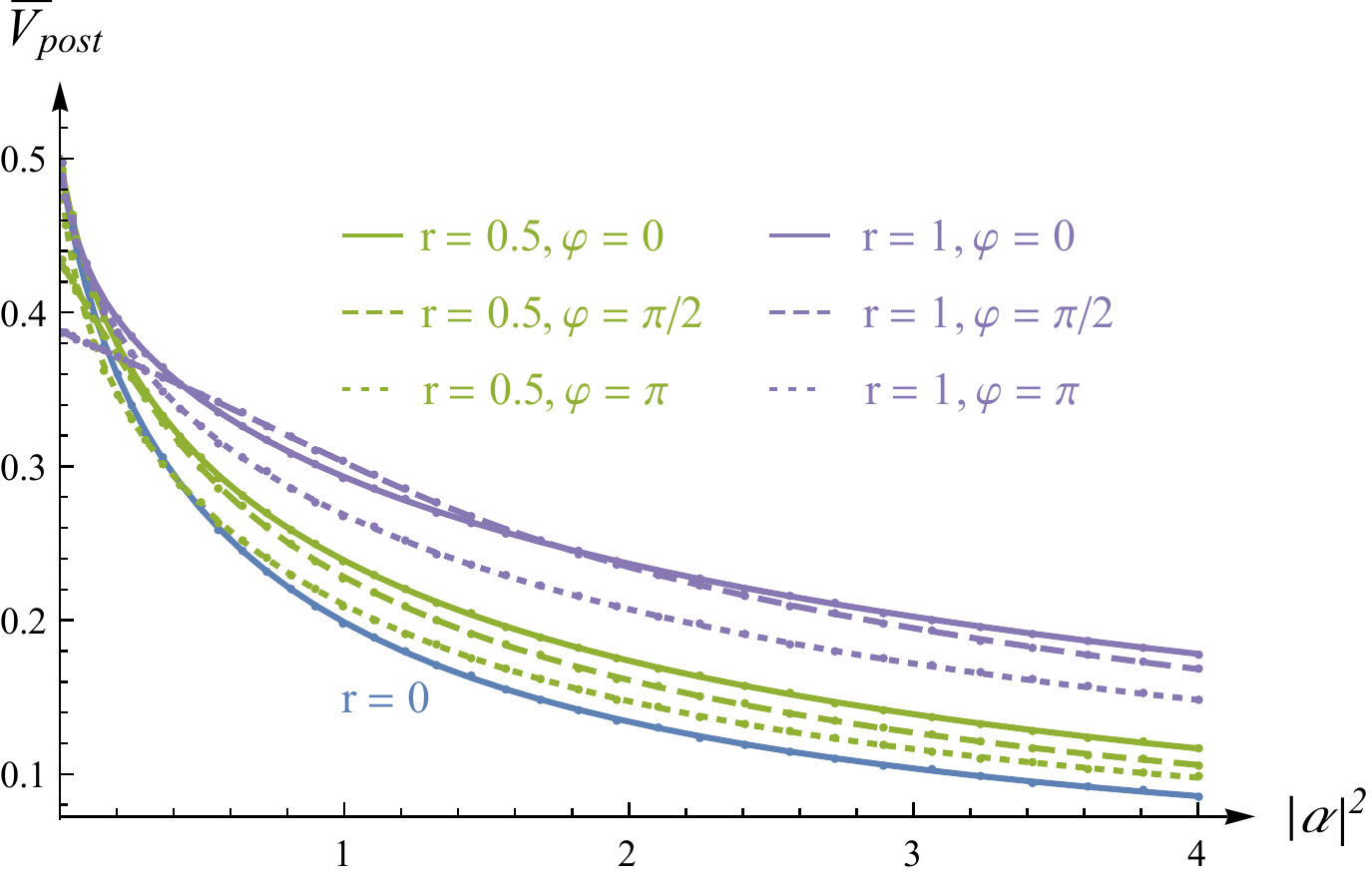}\\
        (a)\\
        \includegraphics[width=0.9\columnwidth]{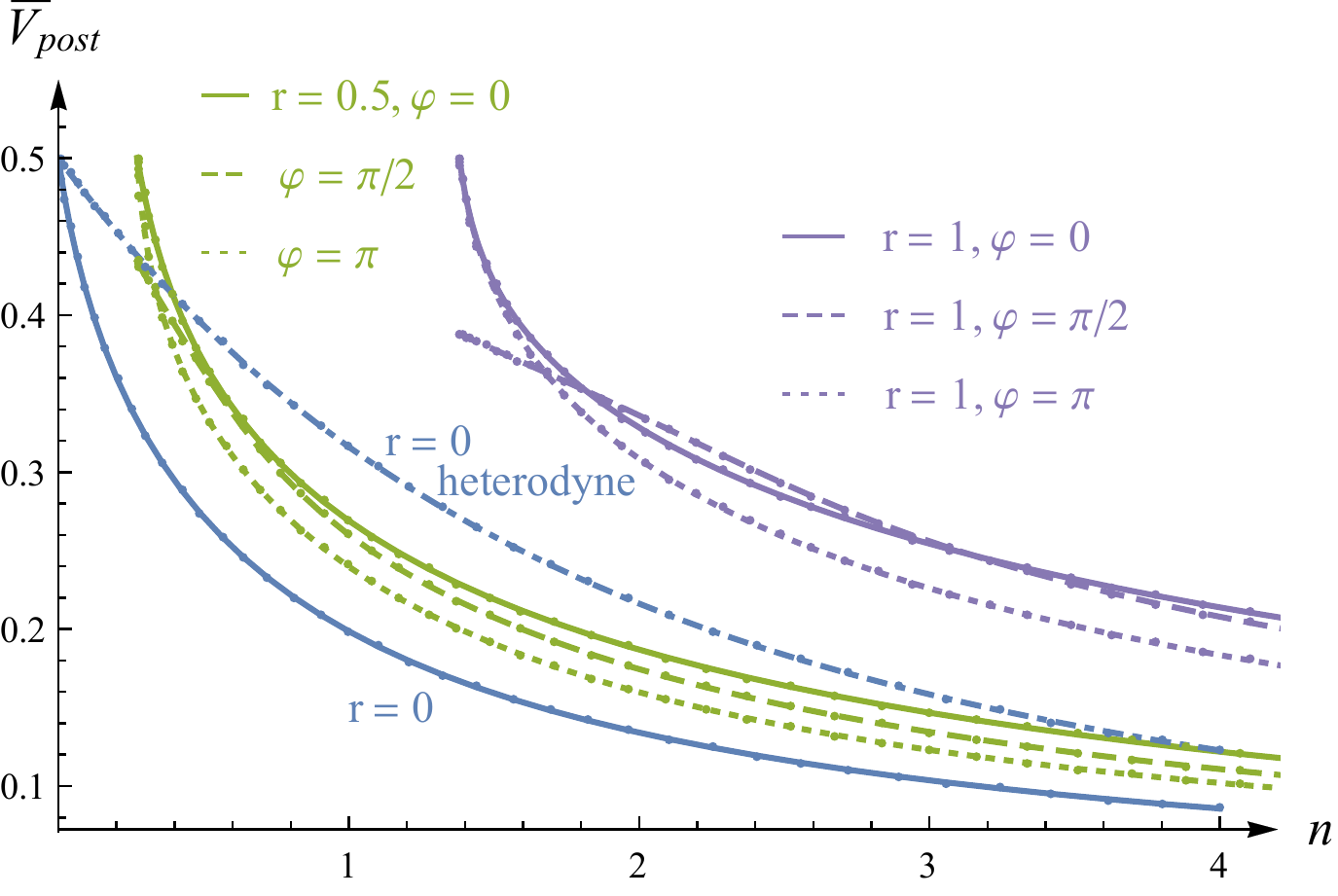}\\
        (b)\\
        \caption{
        Bayesian phase estimation with single-mode Gaussian probes. 
       (a) The average variance $\bar{V}_{\mathrm{post}}$ is shown as a function of $|\alpha|^{2}$, i.e., the energy invested in displacing the probe state. Each curve corresponds to varying values of $\alpha\geq0$, but fixed squeezing strength $r$ from $r=0$ (blue), over $r=0.5$ (green), to $r=1$ (purple),  
       and fixed squeezing angle $\varphi$, from $\varphi=0$ (solid), over $\varphi=\pi/2$ (dashed), to $\varphi=\pi$ (dotted). Curves for $\varphi=3\pi/2$ are identical to those for $\varphi=\pi/2$. 
       (b) The average variance $\bar{V}_{\mathrm{post}}$ is shown as a function of the average photon number $n=\alpha^2+\sinh^2{r}$ of the probe state. The colour-coding is the same as in (a), but the lines do not start at $n=0$ because the nonzero values of $r$ give rise to non-zero average energies even for $\alpha=0$. In addition, (b) shows $\bar{V}_{\mathrm{post}}$ for a coherent probe state ($r=0$) and heterodyne detection from Eq.~(\ref{eq:ageV_phase_displ}) as a blue dashed-dotted curve.
       }
       \label{fig:phase_homo}
    \end{figure}

    For the homodyne measurement, the likelihood to obtain outcome $q$ given the phase $\theta$ can again be obtained by integrating the Wigner function from Eq.~(\ref{eq:wignerfcn}) over the $\hat{p}$-quadrature as in Eq.~(\ref{eq:likelihood_coheret_homo_phase 1}).
    To this end, we note that the vector of first moments is again $\mathbf{\bar{x}}=\sqrt{2}\alpha(\cos\theta,-\sin\theta)^{T}$, while the covariance matrix is given by Eq.~(\ref{eq:cov matrix single mode states}) but with $\varphi\rightarrow\varphi+2\theta$.
    Accordingly, we find the likelihood
    \begin{align}
        p(q\nr|\theta) 
        &=|\langle q| \hat{R}(\theta)\hat{D}(\alpha)\hat{S}(r \mathrm{e}^{i\varphi})|0\rangle|^{2}\nonumber\\
        &=
        \frac{
            \exp\Bigl[-\tfrac{\left(x - \sqrt{2}\alpha\cos{\theta}\right)^2}
        {\Gamma_{qq}(r,\varphi+2\theta)}\Bigr]
        }
        {
        \sqrt{\pi
        \Gamma_{qq}(r,\varphi+2\theta)
        }
        },\label{eq:p of q homodyne squeezed displaced}
    \end{align}
    where $\Gamma_{qq}(r,\varphi)=\cosh(2r)-\cos(\varphi)\sinh(2r)$.
    The (unconditional) probability $p(q)$ to obtain $q$ is $p(q)=\int_{0}^{\pi}d\theta \, p(\theta) \, p(q|\theta)$.
    However, as anticipated from the already complicated form of $p(q)$ for purely displaced probe states, the integration of $p(q)$ from Eq.~(\ref{eq:p of q homodyne squeezed displaced}) turns out to be a formidable obstacle and we have not found a closed analytical expression for it. 
    From this point onward, we hence proceed by numerically evaluating $p(q)$, the posterior $p(\theta\nr|\nr q)$, the estimator, the variance, and the average variance for different displacement strengths ($r$) and angles ($\varphi$) as well as for different displacements $\alpha$.
    In particular, we plot the resulting average variance $\bar{V}_{\mathrm{post}}$ as a function of $|\alpha|^{2}$ and as a function of the average photon number $n=|\alpha|^2+\sinh^2\,{r}$ in Figs.~\ref{fig:phase_homo}~(a) and~(b), respectively. 
    
    We first observe that the average variance for the vacuum state ($\alpha=0=r$) is $1/{2}$, the same value as for the flat prior. 
    Indeed, any non-zero squeezing appears to improve upon this probe state. However, for increasing displacements, squeezing seems to have a detrimental effect compared to purely displaced states with the same $\alpha$ as seen in Fig.~\ref{fig:phase_homo}~(a), where the average variance of purely displaced states is the smallest except in a regime of small $\alpha$.
    When comparing probe states at fixed average energy, it becomes even more clear that squeezing of the probe states in combination with homodyne detection results in strictly worse performance relative to purely displaced probe states.
    Moreover, a comparison with the combination of coherent probe states and heterodyne detection suggests that coherent probe states and homodyne detection outperform any strategy for Bayesian phase estimation (with flat priors) using Gaussian states and heterodyne detection. 
    However, we note that homodyne detection does not allow us to distinguish between phases shifted by $\pi$. 
    If one wishes to explore the full range from $[-\pi,\pi)$, heterodyne detection should be chosen instead. 
%%%%%%%%%%%%%%%%%%%%%%%%%%%%%%%%%%%%%%%%%%%%%%%%%%%%%%%%%%%%%%%%%%%%%%%%%%%%%%%%%%%%%%%

\section{Squeezing estimation}
\label{sec:squeezing}
    In this section we present a Bayesian estimation strategy for estimating the squeezing strength $r$ of a squeezing operation $\hat{S}(\xi)$, where $\xi=r\mathrm{e}^{i\varphi}$, as defined in Eq.~(\ref{squeezingoperator}). 
    The squeezing angle $\varphi$ is assumed to be known. 
    We make this simplifying assumption here, since the investigation of the Bayesian estimation of the single parameter $r$ alone is already computationally demanding, which would only be exacerbated by considering a two-parameter estimation problem. 
    
    Optimal covariant measurement strategies for variants of this estimation problem have been presented in~\cite{MilburnChenJones1994,ChiribellaDArianoSacchi2006}. 
    However, the corresponding optimal POVMs may be sufficiently more difficult to realize practically than the Gaussian measurements we consider here. 
    Moreover, we will focus on investigating the performance of different probe states using solely homodyne detection. 
    This is motivated by the findings of the previous sections, namely, that Gaussian strategies for Bayesian single-parameter estimation based on homodyne detection typically outperform those based on heterodyne detection. 
    As we have previously mentioned, this may be a consequence of the intrinsic uncertainties of the coherent states corresponding to the outcomes of the heterodyne measurement.
    This intuition is also backed up by similar observations made in~\cite{ChiribellaDArianoSacchi2006, GaibaParis2009}, as well as tentative numerical comparisons we have made. 
    The aim of this section is hence to identify practically realizable strategies for estimating the squeezing strength based on single-mode Gaussian states and homodyne detection.
    Nevertheless, we should mention here that heterodyne detection should not be disregarded entirely, since there may be scenarios, such as the simultaneous estimation of squeezing strength and angle, where such a strategy could prove to be advantageous. 
    
    In the remainder of this section, we consider a general pure Gaussian probe state $\hat{D}(\alpha)\hat{S}(\chi)\ket{0}$, where we write the complex variables $\alpha=\alpha_\mathrm{R}+i\alpha_\mathrm{I}$ for $\alpha_{\mathrm{R}},\alpha_{\mathrm{I}}\in\mathbb{R}$ and $\chi =s\mathrm{e}^{i\psi}$, with vector of first moments $\mathbf{\bar{x}}$ and covariance matrix $\sigma$. 
    The squeezing transformation that is to be estimated can be represented by a symplectic matrix $M$, 
    \begin{align}
        M=\begin{pmatrix} \cosh{r} - \cos{\varphi}\ \sinh{r} & \sin{\varphi}\ \sinh{r}\\
        \sin{\varphi}\ \sinh{r} &  
      \cosh{r}+ \cos{\varphi}\ \sinh{r}\end{pmatrix},
    \end{align}
    such that the moments of the Wigner function change according to $\mathbf{\bar{x}}\mapsto M\mathbf{\bar{x}}$ and $\sigma\mapsto M \sigma M^{T}$ under this transformation.
    Since we assume the direction of the unknown squeezing to be known, we may choose our reference frame accordingly and set $\varphi=0$ and $r\in\mathbb{R}$ without loss of generality. 
    
    Although homodyne detection is not a covariant measurement (cf. definition in Sec.~\ref{sec:displacent}), it is still a Gaussian measurement (cf. definition in Sec.~\ref{sec:gaussian_measurements}).
    Consequently, the likelihood $p(q\nr|\nr r)$ is a Gaussian distribution given by
    \begin{align}
        p(q\nr|\nr r)&=|\langle q|\hat{S}(r)|\alpha,\nr \chi\rangle|^2\notag \\
        &=        \frac{\exp(\frac{-\mathrm{e}^{2r}(\sqrt{2}\alpha_\mathrm{R}\mathrm{e}^{-r}-q)^2}{\cosh{2s}-\cos{\psi}\ \sinh{2s}})}{\mathrm{e}^{-r}\sqrt{\pi(\cosh{2s}-\cos{\psi}\ \sinh{2s})}}. \label{squeezinglikelihoodgeneralprobe}
    \end{align}
    The parameter we wish to estimate is not the mean of the likelihood, but is encoded in both the variance and the mean of $p(q\nr|\nr r)$. 
    This makes an analytical treatment of this problem extremely difficult, especially since the function $\exp(\exp(r))$ is known to have a nonelementary antiderivative. 
%%%%%%%%%%%%%%%%%%%%%%%%%%%%%%%%%%%%%%%%%%%%%%%%%%%%%%%%%%%%%%%%%%%%%%%%%%

\subsection{Vacuum probe state}
    In the present scenario, the only case where the likelihood of Eq.~(\ref{squeezinglikelihoodgeneralprobe}) permits an analytical treatment is the vacuum probe state, i.e., when $\alpha=0$ and $\chi=0$, where the likelihood becomes
    \begin{align}
        p(q\nr|\nr\delta)=\frac{\exp(-\frac{q^2}{2\delta^2})}{\delta\sqrt{2\pi}},
        \label{eq:likelihood vacuum probe homodyne for squeezing estimation}
    \end{align}
    with $\delta:=\mathrm{e}^{-r}/\sqrt{2}$.
    This allows us to use the theory of conjugate priors (see Sec.~\ref{sec:Bayesian estimation}).
    For normal distributions with unknown standard deviation $\delta$, the conjugate priors are gamma distributions. 
    However, since this special case does not provide a promising strategy for the problem at hand, we omit the calculation here and refer the interested reader to Appendix~\ref{appendix:Squeezing estimation using the vacuum state and homodyne detection}. 
    
    Instead of analysing this scenario further, we argue that the vacuum state and even the whole class of squeezed vacuum states perform rather poorly as probes. 
    For probes of this kind the vector of first moments remains unchanged by the transformation and so the parameter has to be estimated solely by the change of the covariance matrix. 
    The most likely measurement outcomes close to the origin are therefore generally very inconclusive.
    This reasoning is backed up by tentative numerical explorations, suggesting poor performance for any squeezed vacuum states. 
    Since this strategy does not appear to perform reasonably well, we explore the class of coherent probe states instead in the next section, before considering more general single-mode probe states in Sec.~\ref{sec:Squeezing estimation with a general pure Gaussian state}.
%%%%%%%%%%%%%%%%%%%%%%%%%%%%%%%%%%%%%%%%%%%%%%%%%%%%%%%%%%%%%%%%%%%%%%%%%%

\subsection{Coherent probe states}
    \begin{figure}[tbp]
    \includegraphics[width=0.8\columnwidth,trim={0mm 0mm 0mm 0mm}]{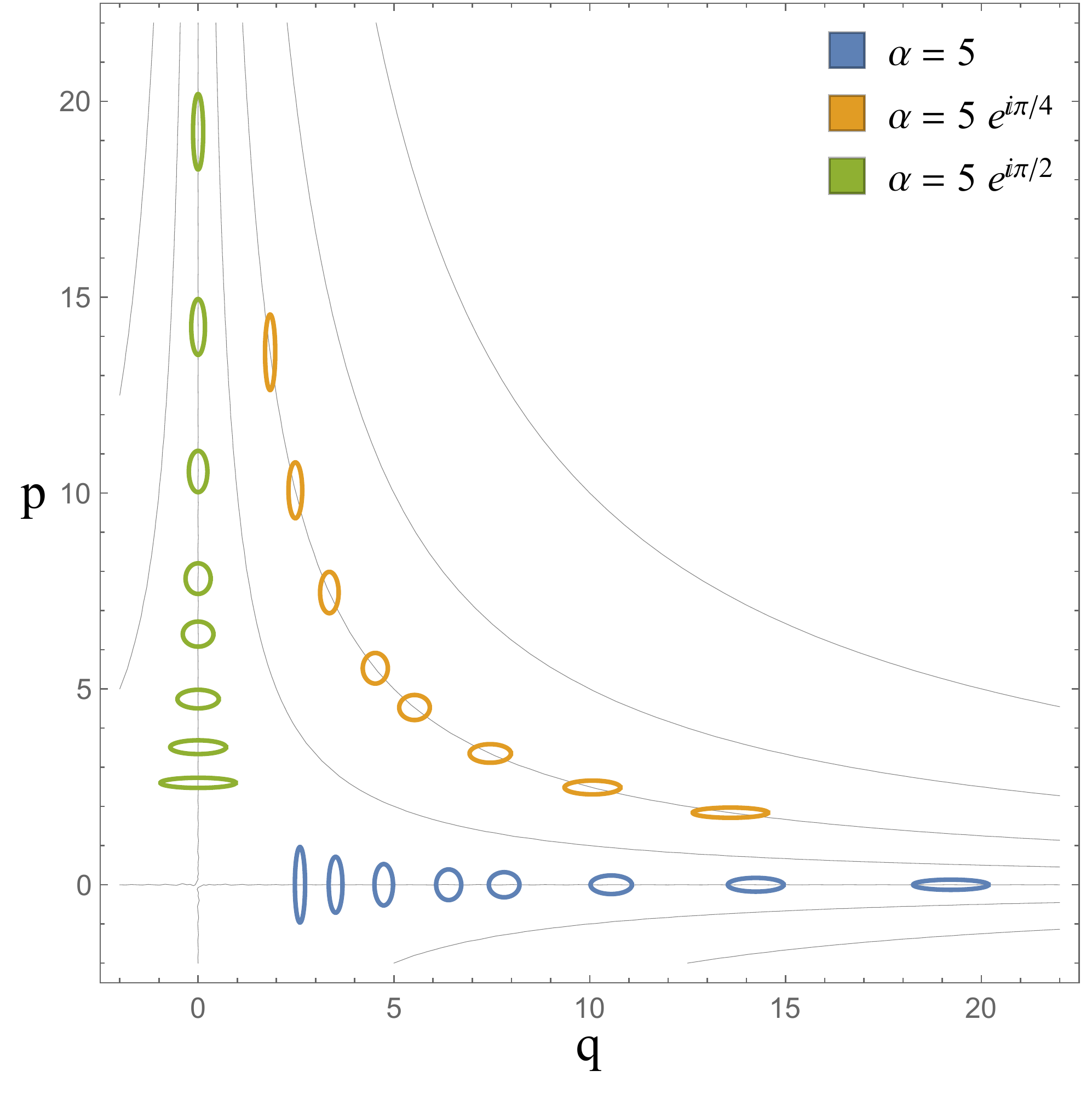}
    \caption{
    Coherent probe states for squeezing estimation. 
    The figure shows cross sections of the Wigner functions of coherent probe states with displacements $\alpha = 5$ (blue), $\alpha =5\ \mathrm{e}^{i\pi/4}$ (orange) and $\alpha =5\ \mathrm{e}^{i\pi/2}$ (green) after the encoding (squeezing) with strength $r=(-1,\ -0.7,\ -0.4,\ -0.1,\ 0.1,\ 0.4,\ 0.7,\ 1)$ has been applied. 
    The axes show the phase space coordinates $q$ and $p$. 
    While the shape of the Wigner function can be seen to change with varying squeezing strengths, the mean values $\expval{\hat{q}}$ and $\expval{\hat{p}}$ can be seen to move along hyperbolic trajectories (grey lines).
    }
    \label{squeezingwithcoherentprobe}
    \end{figure}
    
    For coherent probe states, the parameter $r$ is encoded both in the mean and the variance of the likelihood, see Eq.~(\ref{squeezinglikelihoodgeneralprobe}).
    This makes the estimation more efficient, as probes encoded with different values of the parameter become more distinguishable. 
    
    Under the influence of a squeezing transformation with unknown strength the mean of our probe state moves along hyperbolic trajectories in phase space, as illustrated in Fig.~\ref{squeezingwithcoherentprobe}. 
    To simplify our analysis, we pick a trajectory corresponding to a straight line for our estimation. 
    All states with purely real or imaginary displacement lie on such a trajectory (e.g., the states whose Wigner functions are shown in blue and green in Fig.~\ref{squeezingwithcoherentprobe}) and without loss of generality we assume a positive (real) displacement in $\hat{q}$ together with a homodyne detection in $\hat{q}$. 
    Now the distinguishability of the states with respect to a measurement in $\hat{q}$ is maximal, since the measurement direction is always parallel to the change of the probes mean, ensuring a globally stable measurement procedure. This would not hold for the other hyperbolic trajectories, where the optimal direction of the homodyning (tangential to the curve) would depend on the location on the curve, i.e., the unknown squeezing strength.
    
    With these justified simplifications, our scenario now only has one degree of freedom in the probe preparation, i.e., the displacing amplitude, and none in the measurement basis.
    
    \begin{figure}[tbp]
    \centering
        \includegraphics[width=1\columnwidth,trim={0mm 0mm 0mm 0mm}]{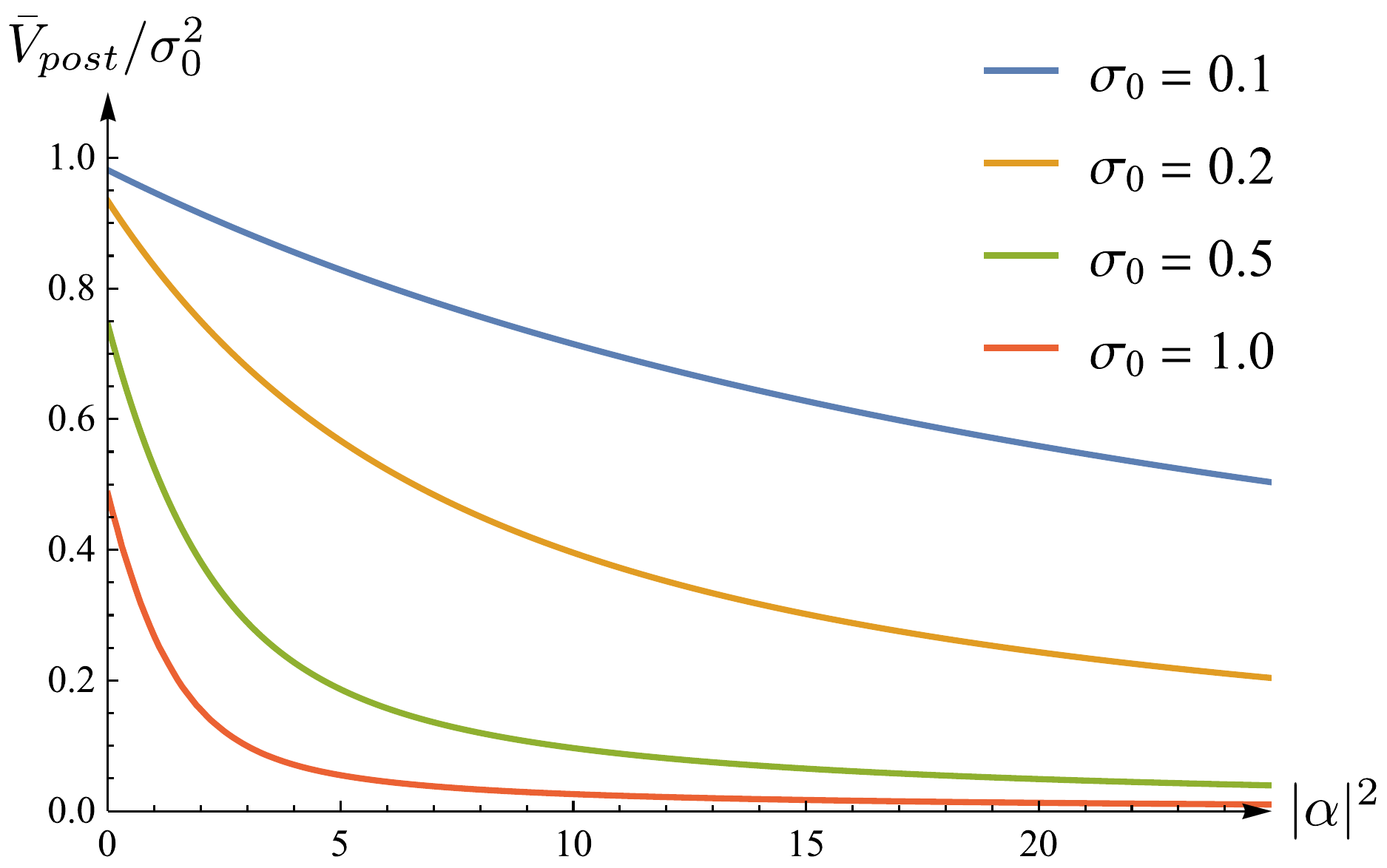}
    \caption{Ratio between posterior and prior variance for squeezing-strength estimation with coherent probe. 
    The plot shows the quotient of the average variance of the posterior and the variance of the prior plotted against the displacement of the probe state. Different lines show different prior variances. 
    The rate at which we acquire knowledge about the squeezing parameter $r$ decreases for increasing knowledge of that parameter. The prior is a normal distribution with mean $r_0=1$ and variance $\sigma_{0}^{2}$.
    }
    \label{squeezingwithcoherentprobe2}
    \end{figure}
    
    In Fig.~\ref{squeezingwithcoherentprobe2} we show numerical results, indicating already a remarkably good performance of this estimation strategy.
%%%%%%%%%%%%%%%%%%%%%%%%%%%%%%%%%%%%%%%%%%%%%%%%%%%%%%%%%%%%%%%%%%%%%%%%%

\subsection{Displaced-squeezed probe states}\label{sec:Squeezing estimation with a general pure Gaussian state}
\begin{figure}[tbp]
    \centering
    \subfigure[]{
        \includegraphics[width=0.95\columnwidth,trim={0mm 0mm 0mm 0mm}]{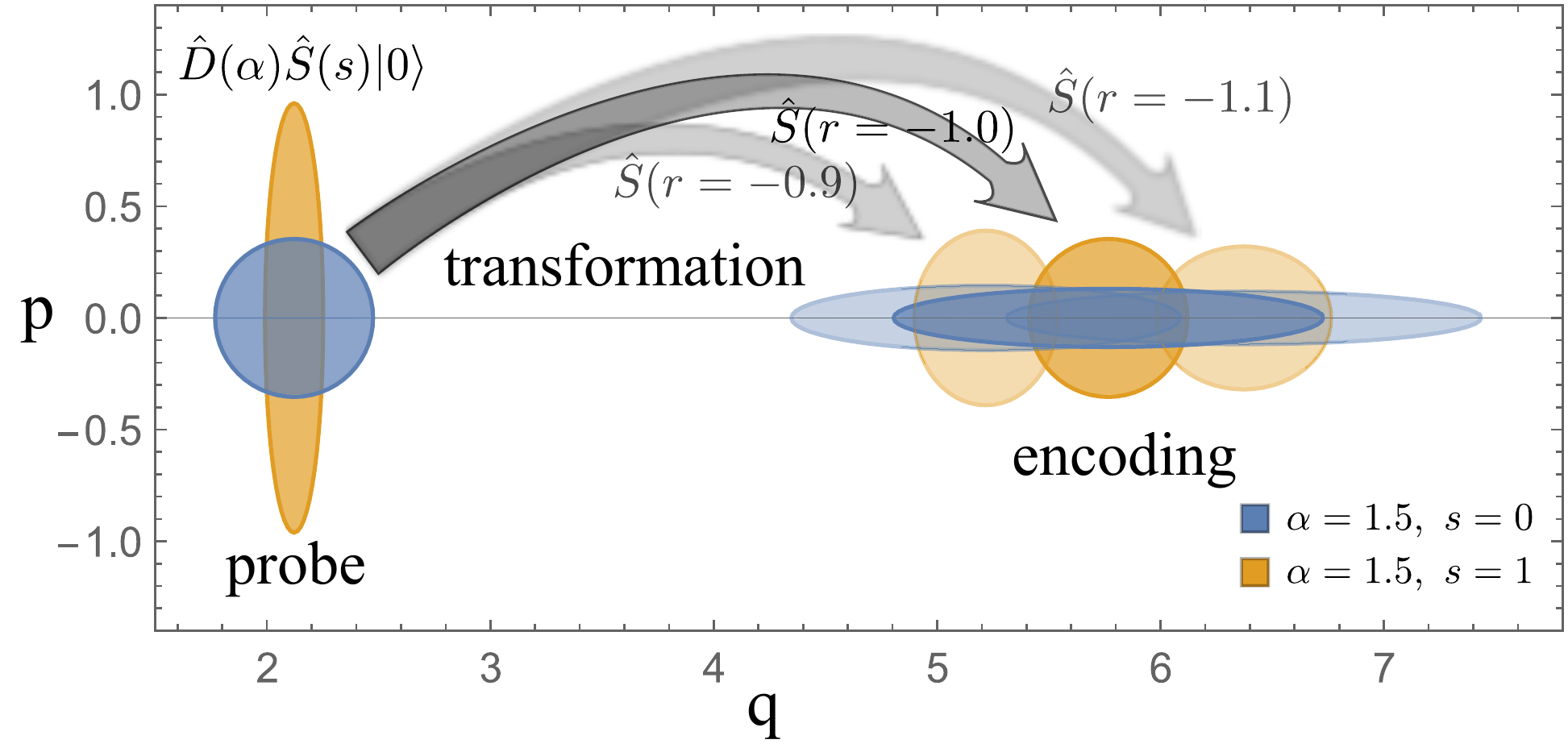}}
    \vspace*{-2mm}
    \subfigure[]{
        \includegraphics[width=0.95\columnwidth,trim={0mm 0mm 0mm 15mm}]{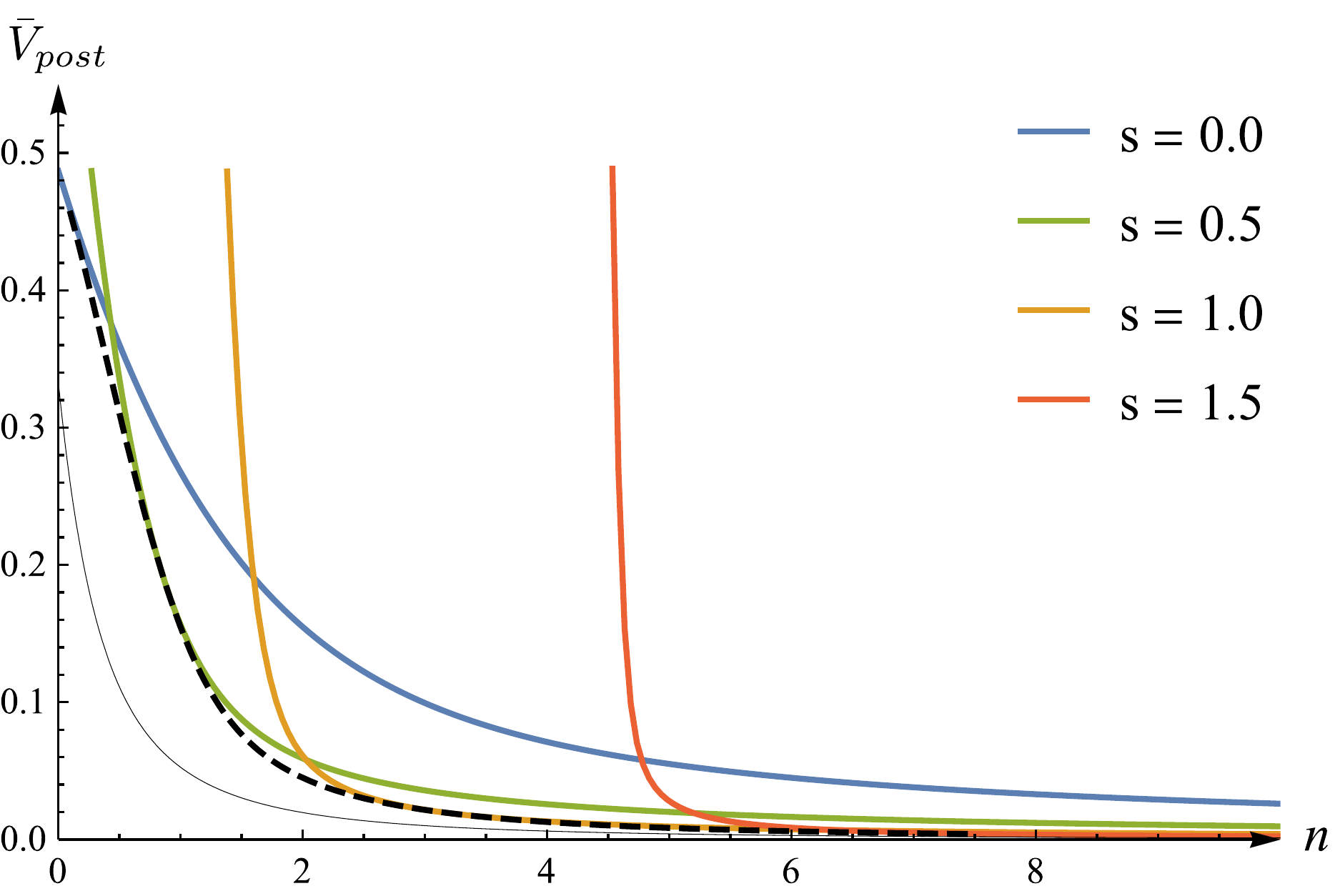}}
    \vspace*{-2mm}
    \subfigure[]{
        \includegraphics[width=0.95\columnwidth,trim={0mm 0mm 0mm 0mm}]{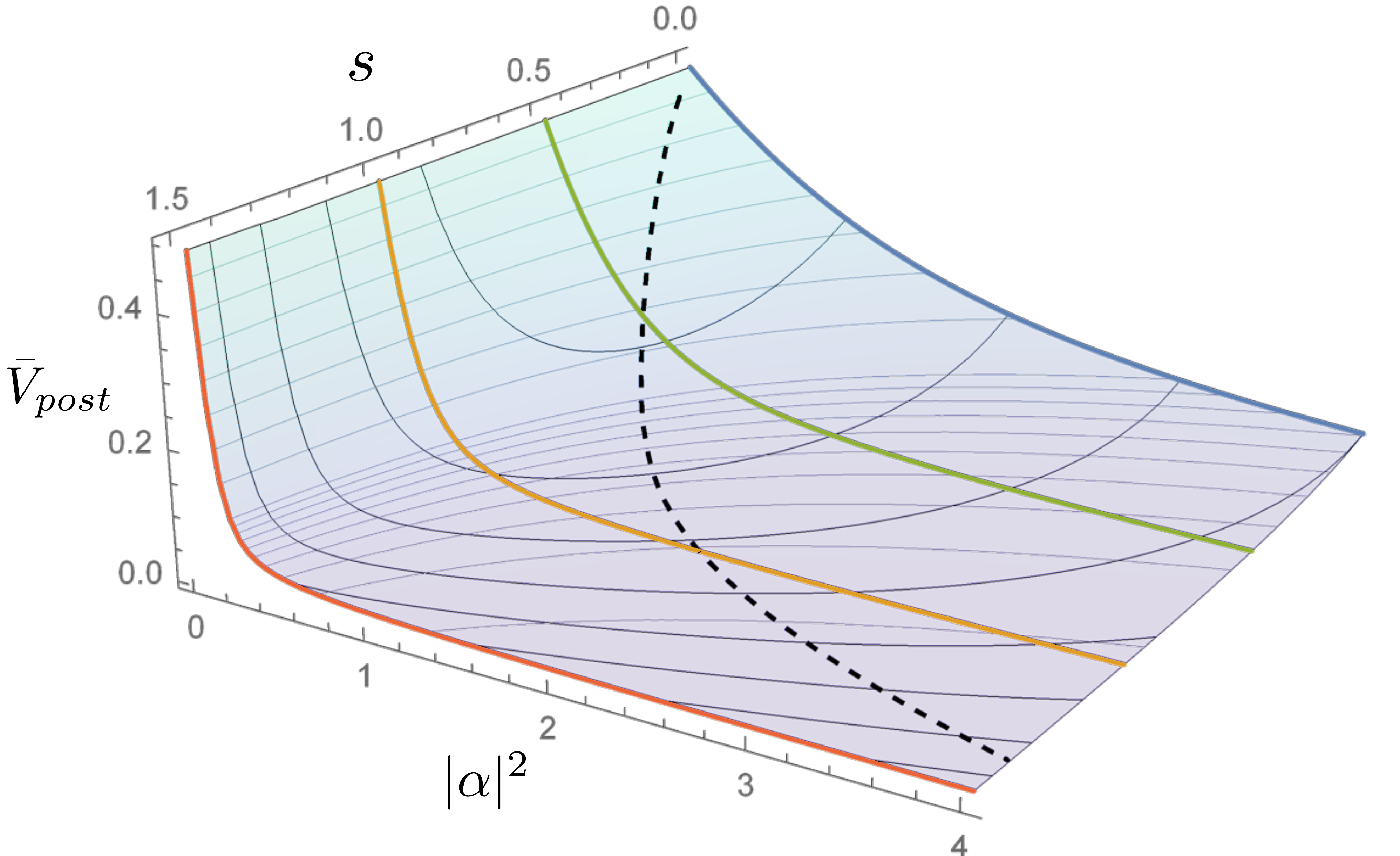}}
    \vspace*{-2mm}
    \caption{Initial squeezing and displacement improve the estimation of the squeezing strength $r$. 
    (a) shows this behaviour in phase space: two differently squeezed probe states (left side) are transformed with an unknown squeezing transformation. For slightly different squeezing strengths ($r=-0.9,-1.0,-1.1$) the unsqueezed probe state ($s=0$, blue) overlaps for the different cases, thus making it hard to estimate the parameter exactly. 
    The initially squeezed probe state ($s=1$, orange) is still clearly distinguishable after the different transformations.
    (b) shows the average variance of the posterior $\bar{V}_{\mathrm{post}}$ as a function of the average photon number $n$ for the two states from (a) and for two more probes with $s=0.5$ (green) and $s=1.5$ (red). The dashed, black line shows $\bar{V}_{\mathrm{post}}$ achieved with the optimal single-mode Gaussian states at fixed $n$. The solid, black line shows the lower bound given by the Van Trees inequality (see Sec.~\ref{sec:Squeezing estimation with a general pure Gaussian state}).
    (c) shows $\bar{V}_{\mathrm{post}}$ for different values of the squeezing $s$ and displacement $\alpha$ of the probe state. The black curves represent lines of constant photon number [$n=|\alpha|^2+\sinh^2(s)$], whereas the dashed, black line minimises the average variance for fixed $n$. 
    The four curves from (b) are shown in the same color-coding.
    The prior used in both (b) and (c) is a normal distribution with mean $r_0=-0.5$ and variance $\sigma_0^2=1$.
    }
    \label{presqueezing}
    \end{figure}

    To improve our method further, we reduce the uncertainty in the $\hat{q}$-quadrature direction in a similar fashion as in Sec.~\ref{sec:displacent} for displacement estimation, i.e., we reduce the uncertainty of the probe in the direction we are interested in by squeezing it beforehand. Fig.~\ref{presqueezing} (a) illustrates this in phase space. Fig.~\ref{presqueezing} (b) shows how the performance of the estimation is improved by increasing the initial squeezing of the probe and compares the results to the Van Trees bound of Eq.~(\ref{eq:QBCR_bound}). There, the prior is taken to be a normal distribution with variance $\sigma_{0}^{2}=1$, such that $I\bigl[p(r)\bigr]=1$, and the QFI is optimized over all single-mode Gaussian states with fixed average photon number $n$, which yields $\mathcal{I}\bigl[\rho(\theta)\bigr]=2(2n+1)^{2}$, see~\cite[Eqs.~(16) and~(18)]{GaibaParis2009}. This inequality gives a lower bound on the average posterior variance, but it is unclear if there exists strategies that can saturate it. In Fig.~\ref{presqueezing}~(c), the use of squeezing and displacement in the preparation of the probe are directly compared, and the optimal combinations of these two operations for mean photon number are identified. 
    
     Although homodyne detection is not the optimal (maximising the FI) POVM for squeezing estimation in the local/frequentist regime, our analysis provides efficient estimation strategies using only elementary quantum optics methods. 
     In particular, these strategies rely only on single-mode Gaussian states and homodyne detection, allowing a comparably straightforward experimental implementation.
%%%%%%%%%%%%%%%%%%%%%%%%%%%%%%%%%%%%%%%%%%%%%%%%%%%%%%%%%%%%%%%%%%%%%%%%%%%%%%%%%%%%%%%

\section{Discussion \& Conclusion}
\label{sec:conclusions}
    In this paper, we have aimed to provide a comprehensive investigation of Bayesian parameter estimation with single-mode Gaussian states and suitable Gaussian measurements. 
    Notably, the Bayesian approach allows us to study regimes of uncertainty for the estimated parameter (e.g., flat priors, single measurements), where local estimation is not justified. 
    Our focus has not been on finding optimal states and measurements maximising the quantum Fisher information. 
    Instead, we have focused on discovering what can be achieved with practically easily realizable techniques: single-mode Gaussian states combined with heterodyne and homodyne detection. 
    Besides the relevance for experimental implementations, this investigation of single-mode Gaussian states within the theory of Bayesian estimation also creates an important reference point for future explorations of more complicated probe states and measurements.
    Within this setting, we have investigated three paradigmatic cases of CV quantum metrology: the estimation of displacements, phase rotations, and single-mode squeezing strengths. 
    For the Bayesian estimation of displacements, we provide a fully analytic treatment for Gaussian priors, and for arbitrary single-mode states combined with heterodyne or homodyne detection.
    For the estimation of a single phase-space coordinate we prove the optimality of the presented strategy. 
    This optimal strategy entails investing all available energy into squeezing the probe state in the direction of the displacement and a homodyne measurement in the same direction.
    
    For Bayesian phase estimation, many standard techniques from Bayesian parameter estimation have to be adapted to circular statistics. 
    This makes it challenging to explore this scenario analytically, and we therefore focus on the case of flat priors (i.e., no initial information about the phase) as a polar opposite to the well-studied problem of local phase estimation. 
    We provide closed expressions for the average variance achieved for coherent probe states and heterodyne detection. 
    For all other scenarios we rely on numerical calculations, which show that homodyne detection generally outperforms heterodyne detection when restricting the phase to the interval $[0,\pi]$.
    In this case, it is best to invest all available energy into displacing the probe.
    
    Finally we consider the estimation of an unknown squeezing strength. Almost all calculations here have to be done numerically. 
    For this we make a series of well justified assumptions and restrict the large parameter space to a small subset, i.e., the displacement and squeezing of the probe state. 
    Our analysis suggests that the best strategy in this case is to split the energy of the probe state amongst squeezing and displacement, and to perform homodyne measurements.
    
    We envisage the results presented here as a first step in the exploration of Gaussian probe states and measurements in the framework of Bayesian parameter estimation. 
    A number of interesting questions regarding optimality, as well as adaptive multi-round schemes come to mind. This could include the adaptive estimation of both coordinates of the complex displacement parameter with homodyne detection alternating in the measurement quadrature as well as adaptive schemes for phase estimation with more general prior functions. 
    Also an extension to multi-mode Gaussian states~\cite{SafranekLeeFuentes2015} and the estimation of multiple parameters~\cite{BakmouDaoudIaamara2020} seem fruitful directions for further investigations. 
    Although these problems are thus left open for future research, the present work represents an important connection to the respective local estimation problems in that it provides practical strategies for drastically reducing the uncertainty about the estimated parameter.
    Once this has been achieved, one may employ suitable (e.g., asymptotically optimal) local estimation strategies.
%%%%%%%%%%%%%%%%%%%%%%%%%%%%%%%%%%%%%%%%%%%%%%%%%%%%%%%%%%%%%%%%%%%%%%%%%%%%%%%%%%%%%%%

\begin{acknowledgments}
    We are thankful to Michalis Skotiniotis for spending his time at the beach going through our work and for providing insightful comments. 
    We acknowledge Thomas Busch and Stefan Ataman for useful comments. 
    S.M. and N.F. acknowledge funding from the Austrian Science Fund (FWF): P 31339-N27. 
    A.\,U.\ acknowledges support from IQOQI - Vienna for visiting, financial and computational support from OIST Graduate University, especially the computing resources of the Scientific Computing and Data Analysis section, and financial support from a Research Fellowship of JSPS for Young Scientists.
    E.A. acknowledges funding from the European Union's Horizon 2020 research and innovation programme under the Marie Sk\l{}odowska-Curie IF (InDiQE - EU project 845486). 
\end{acknowledgments}
%%%%%%%%%%%%%%%%%%%%%%%%%%%%%%%%%%%%%%%%%%%%%%%%%%%%%%%%%%%%%%%%%%%%%%%%%%%%%%%%%%%%%%%

\bibliographystyle{apsrev4-1fixed_with_article_titles_full_names_new}
\bibliography{bibfile}
%%%%%%%%%%%%%%%%%%%%%%%%%%%%%%%%%%%%%%%%%%%%%%%%%%%%%%%%%%%%%%%%%%%%%%%%%%%%%%%%%%%%%%%

\newpage
\hypertarget{sec:appendix}
\appendix
\section*{Appendix}
    \renewcommand{\thesubsubsection}{A.\arabic{subsection}.\Roman{subsubsection}}
    \renewcommand{\thesubsection}{A.\arabic{subsection}}
    \renewcommand{\thesection}{A}
    \setcounter{equation}{0}
    \numberwithin{equation}{section}
    \setcounter{figure}{0}
    \renewcommand{\thefigure}{A.\arabic{figure}}
%%%%%%%%%%%%%%%%%%%%%%%%%%%%%%%%%%%%%%%%%%%%%%%%%%%%%%%%%%%%%%%%

\subsection{Phase estimation with heterodyne measurement}
\label{sec:app_phase_estimation_heterodyne}
    In this section, we provide additional details on the calculations for phase estimation using heterodyne detection discussed in Sec.~\ref{sec:phase_hetero} of the main text. 
%%%%%%%%%%%%%%%%%%%%%%%%%%%

\subsubsection{Coherent probe states \& heterodyne measurements}
\label{sec:app_phase_estimation_heterodyne_coherent_probe}
    We begin with the estimator for coherent probe states $\ket{\alpha}$ with $\alpha\in\mathbb{R}$ and $\alpha>0$. 
    In this case, the posterior given outcome $\beta$, is given by
    \begin{align}
        p(\theta\nr|\nr\beta)   &=\,
        \frac{p(\theta)\,p(\beta\nr|\nr\theta)}{p(\beta)}
        \,=\,
        \frac{\mathrm{e}^{2\alpha|\beta|\cos(\theta-\phi_{\beta})}}{2\pi\,I_{0}(2\alpha|\beta|)}.    
    \end{align}
    For evaluating the estimator $\hat{\theta}(\beta)=\arg\langle \mathrm{e}^{i\theta}\rangle_{p(\theta\nr|\nr\beta)}$, we then note that 
    \begin{align}
        \int\limits_{-\pi}^{\pi}\!\!\! d\theta\,
        \mathrm{e}^{2\alpha|\beta|\cos(\theta-\phi_{\beta})}\,
        \sin(\theta-\phi_{\beta})\,=\,0,
    \end{align}
    which implies that 
    \begin{align}
        &\int\limits_{-\pi}^{\pi}\!\!\!\! d\theta\,
        p(\theta\nr|\nr\beta)\,
        \sin(\theta-\phi_{\beta})\,=\\
        &\ \ \ =\,\int\limits_{-\pi}^{\pi}\!\!\!\! d\theta\,
        p(\theta\nr|\nr\beta)\,
        \bigl[\sin\theta\,\cos\phi_{\beta}-\cos\theta\,\sin\phi_{\beta}\bigr]\,
        =\,0.\nonumber
    \end{align}
    Consequently, we have
    \begin{align}
        \tan\phi_{\beta}
        &=\,
        \frac{\sin\phi_{\beta}}{\cos\phi_{\beta}}    
        \,=\,
        \frac{\int\!\! d\theta\,
        p(\theta\nr|\nr\beta)\,
        \sin\theta}{\int\!\! d\theta\,
        p(\theta\nr|\nr\beta)\,
        \cos\theta}\,=\,
        \frac{\Im\bigl(\langle \mathrm{e}^{i\theta}\rangle\bigr)}{\Re\bigl(\langle \mathrm{e}^{i\theta}\rangle\bigr)}, 
    \end{align}
    such that our estimator is simply the phase of the outcome $\beta$, i.e., 
    \begin{align}
        \hat{\theta}(\beta)=\arg\langle \mathrm{e}^{i\theta}\rangle_{p(\theta\nr|\nr\beta)} &=\,
        \phi_{\beta}.
    \end{align}
    
    For the average variance of the posterior, we have to evaluate an integral over all values of $\beta\in\mathbb{C}$, which can easily be done in polar coordinates, i.e., $\beta=|\beta|\,\mathrm{e}^{-i\phi_{\beta}}$ such that $\int d^{2}\nl \beta=\int_{0}^{\infty} d|\beta|\ |\beta|\ \int_{-\pi}^{\pi} d\phi_{\beta}$. 
    With this, we can insert from Eqs.~(\ref{eq:p of beta}) and~(\ref{eq:posterior variance phase est heterodyne}), and calculate
    \begin{align} 
        \bar{V}_{\mathrm{post}}
        &=\int\!\!\! d^{2}\nl\beta\ p(\beta)\,V_{\mathrm{post}}(\beta)\,
        \label{eq:ageV_phase_displ_appendix}\\
        &=\,  \nonumber
        \tfrac{\mathrm{e}^{-\alpha^{2}}}{2\pi\,\Gamma(2)}
        \int\limits_{0}^{\infty}\!\!\! d|\beta|\,\int\limits_{-\pi}^{\pi} \!\!\!d\phi_{\beta}
        \ |\beta|\, 
        \mathrm{e}^{-|\beta|^{2}}
        \,_{0}F_{1}(2;\alpha^{2}|\beta|^{2})\\
        &=\,  \nonumber
        \tfrac{\mathrm{e}^{-\alpha^{2}}}{\Gamma(2)}
        \int\limits_{0}^{\infty}\!\!\! d|\beta|\, |\beta|\, 
        \mathrm{e}^{-|\beta|^{2}}
        \,_{0}F_{1}(2;\alpha^{2}|\beta|^{2})
        =\tfrac{1-\mathrm{e}^{-|\alpha|^2}}{2\,|\alpha|^2},
    \end{align}
    which yields the result as stated in Eq.~(\ref{eq:ageV_phase_displ}).
%%%%%%%%%%%%%%%%%%%%%%%%%%%%%%%%%

\subsubsection{Displaced squeezed probe states \& heterodyne measurements}
\label{sec:app_phase_estimation_heterodyne_coherent_squeezed_probe}
    In this section, we provide additional details on the calculations in Sec.~\ref{sec:displaced squeezed probe and heterodyne} of the main text. 
    There, we consider Bayesian phase estimation using displaced squeezed states $\hat{D}(\alpha)\hat{S}(\xi)$, where $\alpha>0$ and $\xi=r\nr \mathrm{e}^{i\varphi}$ with $r\geq0$ and $\varphi=\pi$, combined with heterodyne detection represented by a POVM $\bigl\{\tfrac{1}{\pi}\ket{\beta}\!\!\bra{\beta}\bigr\}_{\beta\in\mathbb{C}}$ with elements that are proportional to projectors on coherent states $\ket{\beta}=\hat{D}(\beta)\ket{0}$. 
    In this scenario, the likelihood for obtaining measurement outcome $\beta=|\beta|\mathrm{e}^{-i\phi_{\beta}}$ given that the estimated phase has the value $\theta$, given by Eq.~\eqref{eq:likelihood_phase_squeezed} in the main text, can be rewritten as 
    \begin{align}
        p(\beta\nr|\nr\theta)
        &=\,
        \frac{\mathrm{e}^{-\alpha^2(1-\tanh r)-|\beta|^2}}{\pi\cosh r} 
        \,
        \exp \Bigl[2\alpha|\beta|\cos{(\theta - \phi_{\beta})}\Bigr]\,
        \nonumber\\[1mm]
        &\qquad
        \times\exp \Bigl[|\beta|^{2}\,\tanh r\,\cos{[2(\theta-\phi_{\beta})]}\Bigr]\,
        \nonumber\\
        &\qquad 
        \times\exp \Bigl[-2\alpha|\beta|\tanh r\,\cos{(\theta - \phi_{\beta})}\Bigr].
    \end{align}
    We can then use the Jacobi-Anger expansion in terms of the modified Bessel functions of the first kind, i.e.,
    \begin{align} \label{eq:JacobiAnger}
        \mathrm{e}^{x\nr\cos{\theta}}
        &=
        \sum_{n=-\infty}^{\infty} I_n(x) \mathrm{e}^{i\nr n\nr\theta}, 
    \end{align}
    and write the unconditional probability $p(\beta)$ as
    \begin{align}
        p(\beta)    &=\,\tfrac{1}{2\pi}\int\limits_{-\pi}^{\pi}\!\!\!d\theta\,p(\beta\nr|\nr\theta)\,
        =
        \sum\limits_{\substack{n,m_{1},m_{2}\\ =-\infty}}^{\infty}
        \int\limits_{-\pi}^{\pi}\!\!\!d\theta\,\mathrm{e}^{i(n+2m_{1}+m_{2})(\theta-\phi_{\beta})}
        \nonumber\\
        &\ \ \ \times\,
        \frac{\mathrm{e}^{-\alpha^2(1-\tanh r)-|\beta|^2}}{2\pi^{2}\cosh r}\mathrm{e}^{i(n+m_{1}+m_{2})\pi}
        \,I_{n}(-2\alpha|\beta|)\nonumber\\[1.5mm]
        &\ \ \ \times\,I_{m_{1}}(-|\beta|^{2}\tanh r)\,I_{m_{2}}(2\alpha|\beta|\tanh r).
    \end{align}
    We then make use of the identity
    \begin{align}
        \int\limits_{-\pi}^{\pi}\!\!\!d\theta\,\mathrm{e}^{i(n+2m_{1}+m_{2})(\theta-\phi_{\beta})} &=\,
        \begin{cases}
            2\pi & \text{if}\ n=-2m_{1}-m_{2}\\
            0 & \text{otherwise}
        \end{cases},
        \label{eq:useful id}
    \end{align}
    such that we obtain
    \begin{align}
        p(\beta)    &=\,\frac{\mathrm{e}^{-\alpha^2(1-\tanh r)-|\beta|^2}}{\pi\nr\cosh r}
        \sum\limits_{\substack{m_{1},m_{2}\\ =-\infty}}^{\infty}
        \mathrm{e}^{-i\nr m_{1}\nr\pi}\,I_{-2m_{1}-m_{2}}(-2\alpha|\beta|)\nonumber\\[1mm]
        &\ \ \ \times\,I_{m_{1}}(-|\beta|^{2}\tanh r)\,I_{m_{2}}(2\alpha|\beta|\tanh r).
    \end{align}
    
    By setting $\mathrm{e}^{-i\nr m_{1}\nr\pi}=(-1)^{m_{1}}$, we thus obtain the expression for the unconditional probability $p(\beta)$ from Eq.~\eqref{eq:pbeta_phase_squdis} of the main text. 
    Using Bayes' law, the posterior is obtained as $p(\theta\nr|\nr\beta)=p(\beta\nr|\nr\theta)/\bigl[2\pi\,p(\beta)\bigr]$.
    
    To evaluate the estimator $\hat{\theta}(\beta)=\arg\langle \mathrm{e}^{i\theta}\rangle_{p(\theta\nr|\nr\beta)}$, we proceed in a similar way as above. 
    We first calculate 
    \begin{align}
        &\langle \mathrm{e}^{i\theta}\rangle
        \,=\,\int\limits_{-\pi}^{\pi}\!\!\!d\theta\,p(\theta\nr|\nr\beta)\,\mathrm{e}^{i\theta}
        \,=\,
        \tfrac{1}{2\pi K}
        \sum_{n_1,n_2,n_3=-\infty}^{\infty} 
        \mathrm{e}^{i(n_1+n_2+n_3)\pi}
        \nonumber\\
        &
        \ \times\, I_{n_1}(-2\alpha\left|\beta\right|)\,
        I_{n_2}(-|\beta|^2\tanh{r})\,
        I_{n_3}(2\alpha|\beta|\tanh{r})
        \nonumber\\[1.5mm]
        &\ \ \ \times
        \int_{-\pi}^{\pi} d\theta\,
        \mathrm{e}^{i\theta}
        \mathrm{e}^{i(n_1+2n_{2}+n_{3})(\theta-\phi_{\beta})}, 
    \end{align}
    where 
    \begin{align}
        K
        :=
        \sum_{m_1,m_2=-\infty}^{\infty}(-1)^{m_1}
        I_{-2m_1-m_2}(-2\alpha\left|\beta\right|) 
        \nonumber\\
        \quad
        \times
        I_{m_1}(-|\beta|^2\tanh{r}) I_{m_2}(2\alpha|\beta|\tanh{r}).
    \end{align}
    Here, we can make use of a similar identity as in Eq.~(\ref{eq:useful id}), i.e., 
    \begin{align}
        &\int\limits_{-\pi}^{\pi}\!\!\!d\theta\,\mathrm{e}^{i\theta}\mathrm{e}^{i(n_1+2n_{2}+n_{3})(\theta-\phi_{\beta})}
        =
        \begin{cases}
            2\pi\,\mathrm{e}^{i\phi_{\beta}} & \!\!\!\text{if}\ \small{n_1=-2n_{2}-n_{3}-1}\\
            0 & \text{otherwise},
        \end{cases}
    \end{align}
    such that we obtain 
    \begin{align}
        \langle \mathrm{e}^{i\theta}\rangle
        &=
        \tfrac{\mathrm{e}^{i\phi_{\beta}}}{K}
        \sum_{n_2,n_3=-\infty}^{\infty} 
        (-1)^{-n_2-1}
        I_{-2n_2-n_3-1}(-2\alpha\left|\beta\right|)
        \nonumber\\
        &\quad
        \times
        I_{n_2}(-|\beta|^2\tanh{r})
        I_{n_3}(2\alpha|\beta|\tanh{r})
        .\label{eq:circular moment given beta}
    \end{align}
    Here, $K\geq0$, since $K=K(\beta)$ is proportional to the probability distribution $p(\beta)$ and the proportionality factor is non-negative. 
    The remaining sum on the right-hand side of Eq.~(\ref{eq:circular moment given beta}) is strictly real-valued, which can be seen by noting that $I_{n}(x)$ is real when both the order $n$ and argument $x$ are real. 
    However, the sum over modified Bessel functions may take positive and negative values. 
    
    If the sum is positive, the estimator corresponds to the phase of the outcome, $\hat{\theta}(\beta)=\arg\langle \mathrm{e}^{i\theta}\rangle_{p(\theta\nr|\nr\beta)}=\phi_{\beta}$, whereas the estimate is shifted by $\pi$ [i.e., $\hat{\theta}(\beta)=\phi_{\beta}+\pi$] if the sum is negative. 
    As seen below (particularly, Eq.~\eqref{eq:int_expsin2}), the distinction between these two cases does not affect the variance of the posterior, because the deviation function $\sin^2\bigl[\theta - \hat{\theta}(\beta)\bigr]$ is invariant under shift by $\pi$. 
    For the variance of the posterior, we take the average of $\sin^2\bigl[\theta - \hat{\theta}(\beta)\bigr]$, and find
    \begin{align}
        V_{\mathrm{post}}(\beta)
        &=\,\int\limits_{-\pi}^{\pi}\!\!\!d\theta\,p(\theta\nr|\nr\beta)\,\sin^2(\theta-\hat{\theta}(\beta))\\
        &=
        \tfrac{1}{2\pi K}\!\!\!
        \sum_{\substack{n_1,n_2,n_3\\ =-\infty}}^{\infty}\!\!\!
        I_{n_1}(-2\alpha\left|\beta\right|)
        I_{n_2}(-|\beta|^2\tanh{r})
        \nonumber\\
        &\qquad
        \times
        I_{n_3}(2\alpha|\beta|\tanh{r})
        \mathrm{e}^{i(n_1+n_2+n_3)\pi}
        \nonumber\\
        &\qquad
        \times
        \int\limits_{-\pi}^{\pi}\!\!\!d\theta\,
        \mathrm{e}^{i(n_1+2n_{2}+n_{3})(\theta-\phi_{\beta})}\,
        \sin^2[\theta - \hat{\theta}(\beta)].
        \nonumber
    \end{align}
    We can again make use of an identify similar to Eq.~(\ref{eq:useful id}), i.e.,
    \begin{align} \label{eq:int_expsin2}
        &\int\limits_{-\pi}^{\pi}\!\!\!d\theta\,\mathrm{e}^{i(n_1+2n_{2}+n_{3})(\theta-\phi_{\beta})}\,\sin^2[\theta-\hat{\theta}(\beta)]
        \nonumber\\
        &=
        \begin{cases}
            \pi & \text{if}\ n_1=-2n_{2}-n_{3}\\
            -\tfrac{\pi}{2} & \text{if}\ n_1=-2n_{2}-n_{3}\pm2\\
            0 & \text{otherwise}
        \end{cases}\;,
    \end{align}
    such that we obtain 
    \begin{align}
        V_{\mathrm{post}}(\beta)
        &=
        \tfrac{1}{2K}
        \sum_{n_1,n_2,n_3=-\infty}^{\infty}
        (-1)^{n_1+n_2+n_3}
        I_{n_1}(-2\alpha\left|\beta\right|)
        \nonumber\\
        &\quad
        \times
        I_{n_2}(-|\beta|^2\tanh{r})
        I_{n_3}(2\alpha|\beta|\tanh{r})
        \nonumber\\
        &\quad
        \times
        \bigl(
        \delta_{n_1,-2n_2-n_3+2}
        \tfrac{-1}{2}
        +
        \delta_{n_1,-2n_2-n_3}
        \nonumber\\
        &\quad\quad+
        \delta_{n_1,-2n_2-n_3-2}
        \tfrac{-1}{2}
        \bigr).
    \end{align}
    To obtain the average variance of the posterior, we switch to polar coordinates, $\beta=|\beta|\,\mathrm{e}^{-i\phi_{\beta}}$, such that
    \begin{align} \label{eq:Vpost_squedisp_hetero}
        \bar{V}_{\mathrm{post}}
        &=\int\!\!\! d^{2}\nl\beta\ p(\beta)\,V_{\mathrm{post}}(\beta)\,
        =\,
        \int\limits_{0}^{\infty}\!\!\! d|\beta|\,\int\limits_{-\pi}^{\pi} \!\!\!d\phi_{\beta}
        \tfrac{\mathrm{e}^{-\alpha^2(1-\tanh{r})}}{2\pi\cosh r}
        \nonumber\\
        &\ \ 
        \times\!\! \sum_{\substack{n_2,n_3\\ =-\infty}}^{\infty}\!\!
        |\beta|
        \, \mathrm{e}^{-|\beta|^2}
        I_{n_2}(-|\beta|^2\tanh{r})
        I_{n_3}(2\alpha|\beta|\tanh{r})
        \nonumber\\
        &\ \ 
        \times \tfrac{1}{2}(-1)^{n_{2}}\Bigl[
        2 I_{-2n_2-n_3}(-2\alpha\left|\beta\right|)
        - I_{2-2n_2-n_3}(-2\alpha\left|\beta\right|)
        \nonumber\\
        &\qquad\qquad\qquad
        - I_{-2-2n_2-n_3}(-2\alpha\left|\beta\right|)
        \Bigr]
        \nonumber\\
        &=
        \tfrac{\mathrm{e}^{-\alpha^2(1-\tanh{r})}}{\cosh r}
        \sum_{\substack{n_2,n_3\\ =-\infty}}^{\infty}
        \int\limits_{0}^{\infty}\!\!\! d|\beta|\,
        |\beta|
        \, \mathrm{e}^{-|\beta|^2}
        \,I_{n_2}(-|\beta|^2\tanh{r})
        \nonumber\\
        &
        \quad
        \times 
        I_{n_3}(2\alpha|\beta|\tanh{r})
        \tfrac{1}{2}(-1)^{n_2}\Bigl[
        2I_{-2n_2-n_3}(-2\alpha\left|\beta\right|)
        \nonumber\\[1mm]
        &
        \qquad
        -I_{2-2n_2-n_3}(-2\alpha\left|\beta\right|)
        -I_{-2-2n_2-n_3}(-2\alpha\left|\beta\right|)
        \Bigr],
    \end{align}
    which coincides with the expression in Eq.~\eqref{eq:aveV_phase_squedisp}. 
    We have not found an analytical expression for the above integral so far, but we have evaluated the integral numerically.
%%%%%%%%%%%%%%%%%%%%%%%%%%%%%%%%%%%%%%%%%%%%%%%%%%%%%%%%%%%%%%%%

\subsubsection{Coherent probe states \& homodyne measurements}
\label{sec:app_phase_estimation_homodyne_coherent_probe}
    Here, we provide additional details on the calculations in Sec.~\ref{sec:coherent probe and homodyne} of the main text. 
    There, we consider Bayesian phase estimation with coherent probe states $\hat{D}(\alpha)\ket{0}=\ket{\alpha}$, where $\alpha>0$, combined with homodyne detection represented by a POVM $\bigl\{\ket{q}\!\!\bra{q}\bigr\}_{\beta\in\mathbb{R}}$. 
    In this scenario, the likelihood for measurement outcome $q$ given the phase $\theta$ is provided by Eq.~\eqref{eq:likelihood_coheret_homo_phase} in the main text, which can be rewritten as 
    \begin{align}
        p(q|\theta)
        &=
        \frac{1}{\pi\sqrt{\pi}}
        \mathrm{e}^{
        -q^2-\alpha^2}
        \int\limits_{0}^{\pi}d\theta
        \mathrm{e}^{
        2\sqrt{2}q\alpha \cos{\theta}
        }
        \mathrm{e}^{
        -\alpha^2
        \cos{(2\theta})}.
    \end{align}
    We express the Jacobi-Anger expansion Eq.~\eqref{eq:JacobiAnger} in a real representation as 
    \begin{align}
        \mathrm{e}^{x\nr\cos{\theta}}
        &=
        I_0(x)
        +
        2\sum_{n=1}^{\infty} I_n(x) \, \cos{(n\nr\theta)}, 
    \end{align}
    since $I_{n}(x)=I_{-n}(x)$. 
    Noticing that the range of $\theta$ is $[0,\pi]$, the (unconditional) probability to obtain outcome $q$ is given by 
    \begin{align}
        p(q)    
        &=\,\tfrac{1}{\pi}\int\limits_{0}^{\pi}\!\!\!d\theta\,p(q\nr|\nr\theta)\,
        =\,
        \tfrac{\mathrm{e}^{-q^2-\alpha^2}}{\pi\sqrt{\pi}}
        \int\limits_{0}^{\pi}\!\!\!d\theta\,
        \Bigl[
        I_0(2\sqrt{2}q\alpha)
        I_0(-\alpha^2)
        \nonumber\\
        &\quad
        +
        2\,I_0(-\alpha^2)
        \sum_{n=1}^{\infty}
        I_n(2\sqrt{2}q\alpha)
        \cos{(n\theta)}\\
        &\quad
        +
        2\,I_0(2\sqrt{2}q\alpha)
        \sum_{m=1}^{\infty}
        I_m(-\alpha^2) 
        \cos{(2m\theta)}
        \nonumber\\
        &\quad
        +
        4
        \sum_{m,n=1}^{\infty}
        I_m(-\alpha^2)
        I_n(2\sqrt{2}q\alpha) 
        \cos{(n\theta)}
        \cos{(2m\theta)}\,
        \Bigr]
        .\nonumber
    \end{align}
    We then use the identities $\int_{0}^{\pi}\!d\theta\,\cos{(n\,\theta)}=0\  \forall\,n\geq 1$ and 
    \begin{align}
        \int\limits_{0}^{\pi}\!\!\!d\theta\,
        \cos{(n\,\theta)}\cos{(2m\,\theta)}
        &=
        \begin{cases}
            \tfrac{\pi}{2} & \text{if}\ \ n=2m\\
            0 & \text{otherwise}
        \end{cases}\; .
    \end{align}
    With this, we obtain 
    \begin{align}
        p(q)    
        &=\,\tfrac{1}{\pi}\int\limits_{0}^{\pi}\!\!\!d\theta\,p(q\nr|\nr\theta)
        \,=\,
        \tfrac{\mathrm{e}^{-q^2-\alpha^2}}{\sqrt{\pi}}
        \Bigl[
        I_0(2\sqrt{2}q\alpha)
        I_0(-\alpha^2)
        \nonumber\\
        &\  
        +2
        \sum_{m=1}^{\infty}
        I_{2m}(2\sqrt{2}q\alpha)
        I_m(-\alpha^2)
        \Bigr]
        \,=\,
        \tfrac{\mathrm{e}^{-q^2-\alpha^2}}{\sqrt{\pi}}\,M,
    \end{align}
    where 
    \begin{align}
        M
        &:=
        \sum_{m=-\infty}^{\infty}
        I_{2m}(2\sqrt{2}q\alpha)
        I_m(-\alpha^2).
    \end{align}
    The posterior $p(\theta|q)$ is then obtained as $p(q|\theta)/[\pi p(q)]$. 
    
    To determine the estimator $\hat{\theta}(q)=\arg\langle \mathrm{e}^{i\theta}\rangle_{p(\theta\nr|\nr q)}$, we calculate $\langle \mathrm{e}^{i\theta}\rangle_{p(\theta\nr|\nr q)}$, i.e.,
    \begin{align}
        \langle \mathrm{e}^{i\theta}\rangle
        &=\,\int\limits_{0}^{\pi}\!\!\!d\theta\,p(\theta\nr|\nr q)\,\mathrm{e}^{i\theta}
        \,=\,
        \tfrac{1}{M}\tfrac{1}{\pi}
        \int\limits_{0}^{\pi}\!\!\!d\theta\,
        \bigl[
        I_0(2\sqrt{2}q\alpha)
        I_0(-\alpha^2)
        \mathrm{e}^{i\theta}
        \nonumber\\
        &\quad
        +
        2I_0(-\alpha^2)
        \sum_{n=1}^{\infty}
        I_n(2\sqrt{2}q\alpha)
        \cos{(n\theta)}
        \mathrm{e}^{i\theta}\\
        &\quad
        +
        2I_0(2\sqrt{2}q\alpha)
        \sum_{m=1}^{\infty}
        I_m(-\alpha^2) 
        \cos{(2m\theta)}
        \mathrm{e}^{i\theta}
        \nonumber\\
        &\quad
        +
        4
        \sum_{m,n=1}^{\infty}
        I_n(2\sqrt{2}q\alpha)
        I_m(-\alpha^2) 
        \cos{(n\theta)}
        \cos{(2m\theta)}
        \mathrm{e}^{i\theta}
        \bigr].\nonumber
    \end{align}
    We then use the identities $\int\limits_{0}^{\pi}\!d\theta\,\mathrm{e}^{i\theta}=2i$, 
    \begin{align}
        \int\limits_{0}^{\pi}\!\!\!d\theta\,
        \cos{(n\nr\theta)}\,
        \mathrm{e}^{i\theta}
        &=
        \begin{cases}
            \pi/2 & \text{if}\ n=1\\[1mm]
            \tfrac{i(1+(-1)^n)}{1-n^2} & \text{if}\ n\geq 2\\
            0 & \text{otherwise}
        \end{cases}\; ,
    \end{align}
    and 
    \begin{align}
        &\int\limits_{0}^{\pi}\!\!\!d\theta\,
        \cos{(n\nr\theta)} \cos{(2m\nr\theta)}
        \mathrm{e}^{i\theta}\\
        &=
        \begin{cases}
            \tfrac{\pi}{4} & \text{if}\ n=2m\pm 1\\
            \tfrac{i(1+(-1)^n)(1-4m^2-n^2)}{(n-2m-1)(n-2m+1)(n+2m+1)(n+2m-1)} & \text{otherwise}
        \end{cases}\; .\nonumber
    \end{align}
    With this, we obtain 
    \begin{align}
        &\langle \mathrm{e}^{i\theta}\rangle
        =
        \tfrac{1}{\pi M}
        \bigg\{2i\nr I_0(2\sqrt{2}q\alpha)
        \Bigl[I_0(-\alpha^2)
        +
        \sum_{m=1}^{\infty}
        I_m(-\alpha^2) 
        \tfrac{2}{1-4m^2}\Bigr]
        \nonumber\\
        &\quad
        +2I_0(-\alpha^2)
        \Bigl[\tfrac{\pi}{2}
        I_1(2\sqrt{2}q\alpha)
        +
        \sum_{n=2}^{\infty}
        I_n(2\sqrt{2}q\alpha)
        \tfrac{i(1+(-1)^n)}{1-n^2}
        \Bigr]
        \nonumber\\
        &\quad
        +
        4
        \sum_{m=1}^{\infty}I_m(-\alpha^2)\,
        \Bigl[\tfrac{\pi}{4}I_{2m-1}(2\sqrt{2}q\alpha)
        +\tfrac{\pi}{4}I_{2m+1}(2\sqrt{2}q\alpha)
        \nonumber\\
        &+\!\!\!
        \sum_{\substack{n=1\\[1pt] n\neq 2m\pm 1}}^{\infty}
        \!\!\!
        I_{n}(2\sqrt{2}q\alpha)
        \tfrac{i(1+(-1)^n)(1-4m^2-n^2)}{(n-2m-1)(n-2m+1)(n+2m+1)(n+2m-1)}
        \Bigr]
        \bigg\}
        \nonumber\\
        %%%%%%%%%%%%
        &=
        \tfrac{i}{\pi M}
        \bigg\{\!
        \sum_{n=1}^{\infty}\!
        \tfrac{4}{1-4n^2}
        \bigl[
        I_0(-\alpha^2)
        I_{2n}(2\sqrt{2}q\alpha)
        \!+\!
        I_n(-\alpha^2) 
        I_0(2\sqrt{2}q\alpha)
        \bigr]
        \nonumber\\
        &\quad
        +
        2
        I_0(2\sqrt{2}q\alpha)
        I_0(-\alpha^2)   
        +
        8
        \sum_{m=1}^{\infty}
        \sum_{n=1}^{\infty}
        I_{2n}(2\sqrt{2}q\alpha)
        I_m(-\alpha^2)
        \nonumber\\
        &\qquad\quad
        \times
        \tfrac{1-4m^2-4n^2}{(2n-2m-1)(2n-2m+1)(2n+2m+1)(2n+2m-1)}
        \bigg\}
        \nonumber\\
        &\quad
        +
        \tfrac{1}{M}
        \bigg\{
        I_0(-\alpha^2)
        I_1(2\sqrt{2}q\alpha)
        +
        \sum_{n=1}^{\infty}I_n(-\alpha^2)
        \Bigl[
        I_{2n-1}(2\sqrt{2}q\alpha)
        \nonumber\\
        &\quad\quad
        +
        I_{2n+1}(2\sqrt{2}q\alpha)
        \Bigr]
        \bigg\}.
    \end{align}
    Finally, we can express the real and imaginary parts of $\langle \mathrm{e}^{i\theta}\rangle$ as 
    \begin{align}\label{appendix:heterophaserealpart}
        \Re[\langle \mathrm{e}^{i\theta}\rangle]
        &=
        \tfrac{
        \sum_{n=-\infty}^{\infty}
        I_{2n+1}(2\sqrt{2}q\alpha)
        I_{n}(-\alpha^2)
        }{
        \sum_{m=-\infty}^{\infty}
        I_{2m}(2\sqrt{2}q\alpha)
        I_m(-\alpha^2)}
    \end{align}
    and 
    \begin{align}\label{appendix:heterophaseimaginarypart}
        \Im[\langle \mathrm{e}^{i\theta}\rangle]
        &=
        \tfrac{2}{\pi}
        \tfrac{
        \sum_{m,n=-\infty}^{\infty}
        I_{2n}(2\sqrt{2}q\alpha)
        I_m(-\alpha^2)}{
        \sum_{k=-\infty}^{\infty}
        I_{2k}(2\sqrt{2}q\alpha)
        I_k(-\alpha^2)}
        \\
        &\quad
        \times
        \tfrac{1-4m^2-4n^2}{(2n-2m-1)(2n-2m+1)(2n+2m+1)(2n+2m-1)},
        \nonumber
    \end{align}
    respectively, where we have used the fact that functions $C_{n,m}$ invariant under the exchanges $n\to-n$ and $m\to-m$ satisfy
    \begin{align}\label{eq:Cn_C-n}
        \sum_{n=1}^{\infty}
        C_{n,m}
        &=
        \tfrac{1}{2}
        \left(
        \sum_{n=-\infty}^{\infty}
        C_{n,m}
        -
        C_{0,m}
        \right)
    \end{align}
    and
    \begin{align}\label{eq:Cnm_C-n-m}
        \sum_{m,n=1}^{\infty}\!\!
        C_{n,m}
        &=
        \tfrac{1}{4}
        \left(
        \sum_{m,n=-\infty}^{\infty}\!\!
        C_{n,m} 
        -
        \sum_{m=-\infty}^{\infty}\!\!
        C_{0,m} 
        -
        \sum_{n=-\infty}^{\infty}\!\!
        C_{n,0}
        +
        C_{0,0}
        \right).
    \end{align}
    The estimator can then be calculated from Eqs.~(\ref{appendix:heterophaserealpart}) and~(\ref{appendix:heterophaseimaginarypart}) via
    \begin{align}
        \hat{\theta}(q) &=\,\arctan\left(\frac{\Im[\langle \mathrm{e}^{i\theta}\rangle]}{\Re[\langle \mathrm{e}^{i\theta}\rangle]}\right).
    \end{align}
%%%%%%%%%%%%%%%%%%%%%%%%%%%%%%%%%%%%%%%%%%%%%%%%%%%%%%%%%%%%%%%%%%%%%%%%%%

\subsection{Squeezing estimation using the vacuum state and homodyne detection}
\label{appendix:Squeezing estimation using the vacuum state and homodyne detection}
    In this appendix, we provide additional details on the estimation of the squeezing strength using a vacuum probe state in combination with homodyne detection. 
    We include this to illustrate that the theory of conjugate priors can applied also in more general cases, even if the calculations might become more involving.
    
    The likelihood is given by Eq.~(\ref{eq:likelihood vacuum probe homodyne for squeezing estimation}),  
    \begin{align}
        p(q\nr|\nr\delta)=\frac{\exp(-\frac{q^2}{2\delta^2})}{\delta\sqrt{2\pi}},
    \end{align}
    where we have defined $\delta:=\mathrm{e}^{-r}/\sqrt{2}$. 
    For normal distributions with unknown standard deviation $\delta$, the conjugate priors are gamma distributions
    \begin{align}
        p(\delta)=\frac{b^{a}\delta^{a-1}\mathrm{e}^{-b\delta}}{\Gamma(a)},
    \end{align}
     $a,\ b>0$. 
     The mean and variance of such a distribution is given by $\operatorname{E}[p(\delta)]=a/b$ and $\operatorname{Var}[p(\delta)]=a/b^2$, respectively. 
     If the prior is gamma distributed with parameters $a$ and $b$, then the posterior after $m$ measurements is gamma distributed as well with parameters $a+m/2$ and $b+\sum\limits_i q_i^2/2$, where $q_i$ is the measurement outcome in each round. 
     The mean and variance of the posterior after $m$ repeated measurements with outcomes $\mathbf{q}=(q_1,\dots,q_m)$ then becomes
    \begin{align}
        \operatorname{E}[p(\delta|\mathbf{q})]&=\frac{2a+m}{2b+\sum\limits_i q_i^2}\\
        \operatorname{Var}[p(\delta|\mathbf{q})]&=\frac{2(2a+m)}{(2b+\sum\limits_i q_i^2)^2},
    \end{align}
    
    From this point on the formulas become really cumbersome. Since homodyning is not a covariant measurement for the squeezing operator, the variance of our posterior distribution depends on the outcome. To calculate the average variance $\int dq\nr p(q)\operatorname{Var}[p(\delta|q)]$, one first needs to calculate
    \begin{align}
        p(q) =& \int d\delta\nr p(\delta)p(q|\delta)\\
        =& \int\limits_0^{\infty}d\delta\frac{\exp(-\frac{q^2}{2\delta^2})}{\delta\sqrt{2\pi}}\frac{b^{a}\delta^{a-1}\mathrm{e}^{-b\delta}}{\Gamma(a)}\notag\\
        =& \frac{1}{\sqrt{\pi}\Gamma(a)}\big[\frac{b}{\sqrt{2}}\Gamma(a-1)\prescript{}{p}{F}_{q}(1;1-\frac{a}{2},\frac{3-a}{2};-\frac{b^2q^2}{8})\notag\\
        &-\frac{b^{a+1}|q|^{a}}{2^{\frac{3+a}{2}}}\Gamma(-\frac{a}{2})\prescript{}{p}{F}_{q}(1;\frac{3}{2},1+\frac{a}{2};-\frac{b^2q^2}{8})\notag\\
        &+\frac{\pi b^{a}|q|^{a-1}}{2^{1+\frac{a}{2}}\Gamma(\frac{1+a}{2})}\sec(\frac{\pi a}{2})\prescript{}{p}{F}_{q}(1;\frac{1}{2},\frac{1+a}{2};-\frac{b^2q^2}{8})\big],\notag
    \end{align}
    where $\prescript{}{p}{F}_{q}(.;.;.)$ is the generalized hypergeometric function (the subscripts p and q are part of the notation for this function and have nothing to do with the phase space coordinates).
    With this now we can calculate the average variance after one measurement $m=1$
    \begin{align}
        \bar{V}_{\mathrm{post}}=&\int dq\nr p(q)\operatorname{Var}[p(\delta|q)]\\
        =& \frac{\sqrt{\pi}(2a+1)}{4\sqrt{b}(a-1)}\prescript{}{p}{F}_{q}(\frac{1}{2};-\frac{1}{2},1-\frac{a}{2},\frac{3-a}{2};\frac{b^3}{4})\notag\\
        &+ \frac{2}{3} b^4(2a+1) \frac{\Gamma(1-a)}{\Gamma(5-a)} \prescript{}{p}{F}_{q}(2;\frac{5}{2},\frac{5-a}{2},3-\frac{a}{2};\frac{b^3}{4})\notag\\
        &\hspace{-30pt}-\frac{\pi^2(2a+1)2^{-a}b^{\frac{3a}{2}-2}\csc(\pi a)}{\Gamma(\frac{a-2}{2})\Gamma^2(\frac{a+1}{2})} \prescript{}{p}{F}_{q}(\frac{a}{2};\frac{1}{2},\frac{a-2}{2},\frac{1+a}{2};\frac{b^3}{4})\notag\\
        &+ \frac{\sqrt{\pi}}{8}(2a+1)b^{\frac{3a-1}{2}}\sec(\frac{\pi a}{2})\frac{\Gamma(-\frac{a}{2})}{\Gamma(a-1)} \notag\\
        &\times\prescript{}{p}{F}_{q}(\frac{1+a}{2};\frac{3}{2},\frac{a-1}{2},\frac{2+a}{2};\frac{b^3}{4})\notag
    \end{align}
    Although we were able to calculate an analytical solution, the result in itself is not interesting, but the techniques we have used might be insightful to the reader.

\end{document}